\documentclass[a4paper,prd,twocolumn,showpacs,superscriptaddress,floatfix,nofootinbib]{revtex4-1}
\usepackage{graphicx}
\usepackage{amsmath}
\usepackage{dcolumn}
\usepackage{bm}
\usepackage{enumitem}   
\usepackage{xcolor}
\usepackage{svg}
\usepackage{natbib}
\usepackage{comment}
\usepackage{hyperref}
\hypersetup{
    colorlinks=true,
    linkcolor=blue,
    citecolor=blue,
    filecolor=magenta,      
    urlcolor=cyan
    }
\usepackage{times}

\newcommand{\cf}{cf.,~}
\newcommand{\eg}{e.g.,~}
\newcommand{\ie}{i.e.,~}

\newcommand{\orcid}[1]{\href{https://orcid.org/#1}{
    \includegraphics[width=10pt]{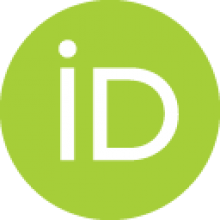}}}
\newcommand{\BHAC}{\texttt{BHAC+}~}
\newcommand{\BHACns}{\texttt{BHAC+}}
\newcommand{\FIL}{\texttt{FIL}~}
\newcommand{\FILns}{\texttt{FIL}}

\begin{document}

\title{A hybrid approach to long-term binary neutron-star simulations}

\author{Harry Ho-Yin Ng\:\orcid{0000-0003-3453-7394}}
\affiliation{Institut f\"ur Theoretische Physik, Goethe Universit\"at,
  Max-von-Laue-Str. 1, 60438 Frankfurt am Main, Germany}

\author{Jin-Liang Jiang\:\orcid{0000-0002-9078-7825}}
\affiliation{Institut f\"ur Theoretische Physik, Goethe Universit\"at,
  Max-von-Laue-Str. 1, 60438 Frankfurt am Main, Germany}

\author{Carlo Musolino\:\orcid{0000-0002-9955-3451}}
\affiliation{Institut f\"ur Theoretische Physik, Goethe Universit\"at,
  Max-von-Laue-Str. 1, 60438 Frankfurt am Main, Germany}

\author{Christian Ecker\:\orcid{0000-0002-8669-4300}}
\affiliation{Institut f\"ur Theoretische Physik, Goethe Universit\"at,
  Max-von-Laue-Str. 1, 60438 Frankfurt am Main, Germany}

\author{Samuel D. Tootle\:\orcid{0000-0001-9781-0496}}
\affiliation{Institut f\"ur Theoretische Physik, Goethe Universit\"at,
  Max-von-Laue-Str. 1, 60438 Frankfurt am Main, Germany}
\affiliation{Department of Physics, University of Idaho, Moscow, ID 83844, USA}

\author{Luciano Rezzolla\:\orcid{0000-0002-1330-7103}}
\affiliation{Institut f\"ur Theoretische Physik, Goethe Universit\"at,
  Max-von-Laue-Str. 1, 60438 Frankfurt am Main, Germany}
\affiliation{School of Mathematics, Trinity College, Dublin 2, Ireland}
\affiliation{Frankfurt Institute for Advanced Studies,
  Ruth-Moufang-Str. 1, 60438 Frankfurt am Main, Germany}

\date{\today}

\begin{abstract}
One of the main challenges in the numerical modelling of binary
neutron-star mergers are long-term simulations of the post-merger remnant
over timescales of the order of seconds. When this modeling includes all
the aspects of the complex physics accompanying the remnant, the
computational costs can easily become enormous. To address this challenge
in part, we have developed a novel hybrid approach in which the solution
from a general-relativistic magnetohydrodynamics (GRMHD) code solving the
full set of the Einstein equations in Cartesian coordinates is coupled
with another GRMHD code in which the Einstein equations are solved under
the Conformally Flat Condition (CFC). The latter approximation has a long
history and has been shown to provide an accurate description of compact
objects in non-vacuum spacetimes. An important aspect of the CFC
approximation is that the elliptic equations need to be solved only for a
fraction of the steps needed for the underlying
hydrodynamical/magnetohydrodynamical evolution, thus allowing for a gain
in computational efficiency that can be up to a factor of $\sim 6~(230)$
in three-dimensional (two-dimensional) simulations. We present here the
basic features of the new code, the strategies necessary to interface it
when importing both two- and three-dimensional data, and a novel and
robust approach to the recovery of the primitive variables. To validate
our new framework, we have carried out a number of tests with various
coordinates systems and different numbers of spatial dimensions,
involving a variety of astrophysical scenarios, including the evolution
of the post-merger remnant of a binary neutron-star merger over a
timescale of one second. Overall, our results show that the new code,
\BHACns, is able to accurately reproduce the evolution of compact objects
in non-vacuum spacetimes and that, when compared with the evolution in
full general relativity, the CFC approximation reproduces accurately both
the gravitational fields and the matter variables at a fraction of the
computational costs. This opens the way for the systematic study of the
secular matter and electromagnetic emission from binary-merger remnants.
\end{abstract}

\maketitle


\section{Introduction}
\label{sec:intro}

A new era of multi-messenger astronomy combining the detections of
gravitational-wave (GW) signals with a variety of electromagnetic
counterparts has begun with the detection of the GW170817 event,
revealing the merger of a system of binary neutron stars
(BNS)~\cite{Abbott2017, Abbott2017_etal, Abbott2017b}. The availability
of multi-messenger signals provides multiple opportunities to learn about
the equation of state (EOS) governing nuclear matter, to explain the
phenomenology behind short gamma-ray bursts and the launching of
relativistic jets~\cite{Rezzolla:2011, Just2016, Ciolfi2020,
  Hayashi2021}, to harvest the rich information coming from the kilonovae
signal~\cite{Metzger:2010, Bovard2017, Smartt2017, Papenfort2018,
  Combi2022, Fujibayashi2023, Kawaguchi2023}, and to obtain information
on the composition of matter accreting around or ejected from these BNS
merger systems (see, \eg \cite{Baiotti2016, Paschalidis2016}, for some
reviews). However, the comprehensive understanding of the physical
mechanisms involved in these phenomena necessitates an accurate and
realistic description of the highly nonlinear processes that accompany
these events. Hence, self-consistent numerical modelling encompassing
accurate prescriptions of the Einstein equations, general-relativistic
magnetohydrodynamics (GRMHD), radiation hydrodynamics to describe
neutrino transport, and the handling of realistic and
temperature-dependent EOSs, plays a fundamental role to achieve this
comprehensive understanding. These techniques are crucial for capturing
the intricate details and the nonlinear dynamics of these systems and
ultimately connect them with existing and future observational data.

Three aspects of the numerical modeling have emerged as crucial now that
a considerable progress has been achieved in terms of the numerical
techniques employed and of the capability of the numerical codes to
exploit supercomputing facilities. The first one is represented by the
ability to carry out simulations on timescales that are ``secular'', that
is, significantly longer than the ``dynamical'' timescale of the inspiral
and post-merger. In fact, over secular timescales, processes such as the
ejection of matter, the development of a globally oriented magnetic field,
or the launching of a jet from the merger remnant can take
place~\cite{Rezzolla2018, Li2018, Metzger2018, Nedora2019, Combi2023,
  Kawaguchi2023}. The second one is the need to have a computational
domain that extends to very large distances from the merger remnant, \ie
extending at least to $10^2-10^3$ $\mathrm{km}$, so as to comprehensively
understand the dynamics of the jet and of the ejected
matter~\cite{Bovard2017, pavan2021, Kiuchi2022, pais2022}. Finally,
achieving extremely high resolution is imperative for accurately
resolving MHD effects during the inspiral~\cite{Giacomazzo:2009mp} and
the associated instabilities after the merger~\cite{Kiuchi2014,
  Carrasco2020b, Chabanov2022}. The combination of these aspects clearly
represents a major challenge in the modelling of BNS mergers and calls
for new approaches where efficiency in obtaining the solution at
intermediate timesteps is optimized.

Essentially all of the numerical schemes that solve the hyperbolic sector
of the Einstein field equations require updating the field variables (\ie
the three-metric tensor, the extrinsic curvature tensor, the conformal
factor, and the gauge quantities) at each Runge-Kutta substep within a
single evolution step. However, an alternative approach involves using a
relatively efficient spacetime solver, such as constraint-enforcing
approaches with a conformally flat condition (CFC), which do not
necessarily require updates at every Runge-Kutta substep or even
evolution step. These CFC approaches typically solve the elliptic sector
of the Einstein equations only every $3-100$ steps of the underlying
hydrodynamical/magnetohydrodynamical evolution, thus allowing, for
example, to capture of highest-frequency pulsation modes in rapidly
rotating neutron stars~\cite{Dimmelmeier02a, Cordero2009, Cheong2020,
  Yip2023} at a fraction of the computational cost. The CFC approximation
has also been successfully used in core-collapse
supernovae~\cite{Dimmelmeier02b, Ott07b, Muller2015}, in rapidly rotating
neutron stars~\cite{Dimmelmeier02a, Cordero2009, Cheong2020, Cheong2021},
and in BNS mergers \cite{Bauswein2012, Bauswein2020c, Lioutas2022,
  Blacker2023}. These studies have demonstrated that the CFC
approximation achieves good agreement with full general relativity (GR),
especially in isolated systems with axisymmetry. It can even reproduce a
similar GW spectrum to simulations using full general relativity for
post-merger remnants following BNS mergers \cite{Bauswein2012}.

Among our ultimate goals -- but also that of much of the community
interested in binary mergers involving neutron stars -- is to investigate
the long-term dynamical properties of BNS post-merger remnants exploiting
an efficient implementation that maintains high accuracy and a complete
description of the microphysics of the neutron-star matter over a
duration of approximately $1$ to $10$ seconds. In addition, we need to
accomplish this by employing a coordinate system that is optimally
adapted to the dynamics of the post-merger object (be it a neutron stars
or a black hole), which is mostly axisymmetric (see, \eg
\cite{Kastaun2014, Hanauske2016}) and where the outflow is mostly radial
and almost spherically symmetric.

To this scope, we here present a novel hybrid approach in which the full
numerical-relativity GRMHD code \FILns~\cite{Most2019b, Most2020e,
  Chabanov2022} is coupled with the versatile, multi-coordinate
(spherical, cylindrical or Cartesian) and multi-dimensional GRMHD
code~\BHACns, which employs the CFC approximation for the dynamics of the
spacetime. We recall that \texttt{BHAC}~\cite{Porth2017, Olivares2019,
  Ripperda2019} was specifically developed to explore black-hole
accretion systems with a stationary spacetime
geometry~\cite{Porth2019_etal}. It possesses robust divergence-cleaning
methods~\cite{Porth2017} and constraint-transport
methods~\cite{Olivares2019} for the enforcement of the divergence-free
condition of the magnetic field. We here present its further development,
\BHACns, which includes a dynamical-spacetime module using the CFC
approximation across three different coordinate systems and an efficient
and reliable primitive-recovery scheme that is coupled with a
finite-temperature tabulated EOS. We also discuss how the coupling
between \FIL and \BHACns, which employ different formulations of the
equations and different sets of coordinates, can be handled robustly and
reliably, either when restricting the simulations to two spatial
dimensions (2D) or in fully three-dimensional (3D) simulations. More
importantly, we show that the hybrid approach provides considerable
savings in computational costs, thus allowing for accurate and robust
simulations over timescales of seconds in 2D and hundreds of milliseconds
in 3D, at a fraction of the computational costs of full-numerical
relativity codes.

The paper is organized as follows. In Sec.~\ref{sec:math_setup}, we
describe the mathematical formulation of the GRMHD equations in a 3+1
decomposition of the spacetime and the Einstein field equations when
expressed under the CFC approximation. Section~\ref{sec:num_setup} is
also used to present the numerical methods and implementation
details. The results of a series of benchmark tests in various dimensions
and physical scenarios are presented in Sec.~\ref{sec:results}, while we
end with a summary and discuss the future aspects in
Sec.~\ref{sec:summary}. Throughout this paper, unless otherwise stated,
we adopt (code) units in which $c = G = M_{\odot} = k_{\rm B} =
\epsilon_0 = \mu_0 = 1$ for all quantities except coordinates. Greek
indices indicate spacetime components (from $0$ to $3$), while Latin
indices denote spatial components (from $1$ to $3$).

\section{Mathematical Setup}
\label{sec:math_setup}

\subsection{Einstein and GRMHD equations}
\label{sec:grmhd}

As mentioned in Sec.~\ref{sec:intro}, we here present a hybrid approach
to BNS merger simulations by combining the solutions obtained from the
full numerical-relativity GRMHD code \FILns~\cite{Most2019b, Most2020e,
  Chabanov2022} with the multi-coordinate and multi-dimensional GRMHD
code \BHACns~\cite{Porth2017, Olivares2019, Ripperda2019}. The main
difference between the two codes is in the way they solve the Einstein
equations, which is performed in \FIL using well-known evolution schemes,
such as BSSNOK~\cite{Shibata95, Baumgarte99}, CCZ4~\cite{Alic:2011a,
  Alic2013}, or Z4c~\cite{Bernuzzi:2009ex}, while \BHAC employs the CFC
approximation with an extended-CFC scheme (xCFC)~\cite{Cordero2009} (see
Sec.~\ref{sec:num_setup} for a short summary). Another difference, but
less marked, is in the way the two codes obtain the solutions of the
GRMHD equations, where different numerical approaches are employed
(again, see Sec.~\ref{sec:num_setup} for additional details). In the
interest of compactness, we will not discuss here the spacetime solution
adopted by \FILns, which is based on well-known techniques reported in
the references above. For the same reason, we will not discuss here the
details of the mathematical formulation of the GRMHD equations, as these
are also well-known and can be found in the works cited above. On the
other hand, we will provide in the next section a brief but complete
review of the CFC approximation of the Einstein equations.

\subsection{The CFC approximation and extended-CFC scheme}
\label{sec:cfc_maths}

Before discussing in detail the practical aspects of our hybrid approach
to the BNS-merger problem, it may be useful to briefly recall the basic
aspects of the CFC approximation. In this framework, which has been
developed over a number of years and has been presented in numerous works
(see, \eg Refs.~\cite{Dimmelmeier02a, Cordero2009}), the spatial
three-metric $\gamma_{ij}$ is obtained via a conformal transformation of
the type
\begin{equation}
\gamma_{i j}= \psi^4 \tilde{\gamma}_{i j} \,,
\end{equation}
where $\psi$ is the conformal factor and $\tilde{\gamma}_{i j}$ the
conformally related metric. As by the name, in the conformally flat
approximation, $\tilde{\gamma}_{i j} = f_{ij}$ with $f_{ij}$ being the
flat spatial metric, so that
\begin{equation}
\partial_t \tilde{\gamma}_{ij} = \partial_t f_{ij} = 0\,.
\end{equation}

Indeed, because of this assumption, which de-facto suppresses any
radiative degree of freedom in the Einstein equations, the CFC is also
known as the ``waveless'' approximation. While this may seem rather crude
at first and forces the use of the quadrupole formula to evaluate the GW
emission from the compact sources that are simulated, a number of studies
have shown the robustness of this approach at least when isolated objects
that possess a sufficient degree of symmetry are considered. In
particular, Ref.~\cite{Ott07b} has shown that the CFC approximation works
exceptionally well in simulations of multi-dimensional rotating
core-collapse supernovae in terms of the hydrodynamical quantities as
well as the gravitational waveforms, and by means of the Cotton-York
tensor~\cite{York71}. In fact, pre-bounce and early post-bounce
spacetimes do not deviate from conformal flatness by more than a few
percent and such deviations reach up to only $\sim~5\%$ in the most
extreme cases of rapidly rotating neutron stars~\cite{Cook1996}, while
the frequencies of fundamental oscillation modes of those models deviate
even less when compared to full general-relativistic simulations
\cite{Dimmelmeier06}. In addition, the dominant frequency contributions
in the GW spectrum of BNS post-mergers simulated with the CFC
approximation deviate at most of several few percent from the full
general-relativistic results obtained with different
EOSs~\cite{Bauswein2012}.

Imposing the CFC approximation, along with the maximal-slicing gauge
condition $K:=\gamma^{ij} K_{ij} = 0$, where $K^{\mu \nu}$ is the
extrinsic curvature, simplifies the Hamiltonian and momentum-constraint
equations of the ADM formulation~\cite{Alcubierre:2006,
  Rezzolla_book:2013}, reducing them to the following set of coupled
nonlinear elliptic differential equations
\begin{eqnarray}
  \label{eq:cfc_eq_psi}
  \hat{\Delta} \psi
  &=& -2 \pi \psi^5 E-\frac{1}{8} \psi^5 K_{i j} K^{i j}\,, \\
  \label{eq:cfc_eq_alp} 
  \hat{\Delta}(\alpha \psi)
  &=& 2 \pi \alpha \psi^5(E+2 S)+\frac{7}{8} \alpha \psi^5 K_{i j} K^{i
    j}\,,
  \label{eq:cfc_eq_alppsi} \\
  \hat{\Delta} \beta^i
  &=&16 \pi \alpha \psi^4 S^i+2 \psi^{10} K^{i j} \hat{\nabla}_j
  \left(\alpha \psi^{-6}\right) \nonumber \\
& &- \frac{1}{3} \hat{\nabla}^i\left(\hat{\nabla}_j \beta^j\right)\,,
\end{eqnarray}
where $\hat{\Delta}$ and $\hat{\nabla}_i$ are the Laplacian and covariant
derivative with respect to the flat spatial metric,
respectively. Furthermore, Eqs.~\eqref{eq:cfc_eq_psi} and
\eqref{eq:cfc_eq_alppsi} employ the following matter-related quantities
\begin{eqnarray}
  &&S_{ij} :=\gamma_{i\mu}\gamma_{j\nu} T^{\mu\nu}\,,\\
  &&S_j := -\gamma_{j\mu} n_\nu T^{\mu\nu}\,,\\
  &&S := \gamma_{ij}S^{ij}\,,\\
  &&E :=n_\mu n_\nu T^{\mu\nu}\,.
\end{eqnarray}
Here, $n_{\mu}$ is the unit timelike vector normal to the spatial
hyperspace, $T_{\mu \nu}$ is the energy-momentum tensor, $S_{ij}$ its
fully spatial projection, $S$ the trace of $S_{ij}$, $S_i$ the momentum
flux, and $E$ the energy density. Also appearing in
Eqs.~\eqref{eq:cfc_eq_psi} and \eqref{eq:cfc_eq_alppsi} are the gauge
functions $\alpha$ and $\beta^i$ -- which are also referred to as the
lapse function and the shift vector, respectively -- so that the
extrinsic curvature under the CFC approximation reads
\begin{equation}
K_{i j}=\frac{1}{2 \alpha}\left(\nabla_i \beta_j+\nabla_j
\beta_i-\frac{2}{3} \gamma_{i j} \nabla_k
\beta^k\right) \label{eq:cfc_eq_kij}\,.
\end{equation}

Due to the nonlinearity of the constraint equations, the original CFC
system of equations~\eqref{eq:cfc_eq_psi}-\eqref{eq:cfc_eq_kij}
encounters problems of non-uniqueness in the solution, particularly when
the configuration considered is very compact. Additionally, the original
CFC system exhibits relatively slow convergence due to the elliptic
equation \eqref{eq:cfc_eq_psi} for $\psi$ that relies on the values of
$K_{ij}$, which themselves depend on $\psi$ and $\beta^{i}$. Since the
equations implicitly depend on each other, this imposes the use of a
recursive-solution procedure that typically requires a large number of
iterations before obtaining a solution with a sufficiently small
error. To avoid these shortcomings, a variant of the original CFC
approach, also known as the xCFC scheme, was firstly introduced in
Ref.~\cite{Cordero2009} and since then widely used in
Refs.~\cite{Bucciantini2011, Cheong2020, Ng2020, Cheong2021, Leung2022,
  Cheong2022, Servignat2023, Cheong2023}.

In the xCFC scheme, the traceless part of the conformal extrinsic
curvature $\hat{A}^{ij}$ is expressed as
\begin{equation}
\hat{A}^{ij} := \psi^{10} K^{ij} = (\boldsymbol{L} X)^{ij} + \hat{A}^{ij}_{_{\rm TT}}\,,
\end{equation}
where
\begin{equation}
(\boldsymbol{L} X)^{ij}:=\hat{\nabla}^i X^j+\hat{\nabla}^j
  X^i-\frac{2}{3}\hat{\nabla}_k X^k f^{i j}\,,
\end{equation}
is the vector potential $X^i$ acted by a conformal Killing operator
associated to the flat spatial metric $\boldsymbol{L}$ and $\hat{A}^{ij}_{_{\rm
    TT}}$ is the transverse traceless part under the conformal transverse
traceless decomposition. Because the amplitude of $\hat{A}^{ij}_{_{\rm
    TT}}$ is smaller than the non-conformal part of the spatial metric
$h_{ij}:= \gamma_{ij} - f_{ij}$ (see Appendix in
Ref.~\cite{Cordero2009}), $\hat{A}^{ij}$ can be approximated under CFC
approximation expressed as
\begin{equation}
\hat{A}^{i j} \approx \hat{\nabla}^i X^j+\hat{\nabla}^j X^i-\frac{2}{3}
\hat{\nabla}_k X^k f^{i j}\,,
\label{eq:xcfc_eq_aij}
\end{equation}
where, the transverse traceless part of $K_{ij}$ is assumed to be much
smaller than $h^{ij}$~\cite{Cordero2009}. The vector potential $X^{i}$
satisfies the following set of elliptic equations that explicitly depend
on the matter source terms
\begin{equation}
  \hat{\Delta} X^i+\frac{1}{3}
  \hat{\nabla}^i\left(\hat{\nabla}_j X^j\right)=8 \pi f^{i j}
  \tilde{S}_j\,,
  \label{eq:xcfc_eq_x}
\end{equation}
where $\tilde{S}_j := \psi^6 S_j$. 

In practice, after evolving the conserved fluid variables ($D, S_j,
\tau$) (see definitions in \cite{Rezzolla_book:2013}), we first define
the rescaled conserved variables
\begin{equation}
\tilde{E} := \psi^{6} E\,, \qquad
\tilde{S}_j := \psi^{6} S_j\,,
\end{equation}
where $\psi$ here means the \textit{old} solution (or first
guess if we are dealing with the initial data) of the conformal factor.
Next, we solve Eq.~(\ref{eq:xcfc_eq_x}) to obtain the solution for
$X^{i}$, which is then used in solving Eq.~(\ref{eq:xcfc_eq_aij}) and
to calculate a \textit{new} estimate for the conformal factor using the
elliptic equation
\begin{equation}
\hat{\Delta} \psi=-2 \pi \psi^{-1} \tilde{E} -\frac{1}{8} \psi^{-7}f_{i k} f_{j l}
\hat{A}^{k l} \hat{A}^{i j} \,.
\label{eq:xcfc_eq_psi}
\end{equation}
In this step, with the updated value for $\psi$, we calculate the
variables and perform the primitive recovery to obtain the primitive
variables needed to evaluate $\tilde{S}_{ij} = \psi^6 S_{ij}$ with the
updated values of $\psi$. We then compute the trace $\tilde{S} :=
\gamma^{ij} \tilde{S}_{ij}$ to obtain the elliptic equations for the
lapse function $\alpha$ and the shift vector $\beta^{i}$, namely
\begin{align}
  &\hat{\Delta}(\alpha \psi)=(\alpha \psi)\left[2 \pi \psi^{-2}(\tilde{E}+2
    \tilde{S}) + \frac{7}{8} \psi^{-8} f_{i k} f_{j l} \hat{A}^{k l} \hat{A}^{i j}
    \right]\,, \label{eq:xcfc_eq_alpha} \\
  &\hat{\Delta} \beta^i+\frac{1}{3} \hat{\nabla}^i\left(\hat{\nabla}_j
  \beta^j\right)=16 \pi \alpha
  \psi^{-6} f^{i j} \tilde{S}_j+2 \hat{A}^{i j}
  \hat{\nabla}_j\left(\psi^{-6} \alpha\right)\,, \label{eq:xcfc_eq_beta}
\end{align}

An important advantage of the xCFC approach is that, thanks to the
introduction of the vector field $X^{i}$, it can be cast in terms of
elliptic equations without an implicit relation between the metric
components and the conformal extrinsic curvature $\hat{A}_{ij}$.
Moreover, since the equations decouple in a hierarchical way, all
variables can be solved step by step with the conserved quantities, thus
increasing the efficiency of the algorithm compared to the original
formulation of CFC scheme (\eg Ref.~\cite{Lioutas2022}). Finally, the
xCFC scheme ensures local uniqueness even for extremely compact
solutions~\cite{Cordero2009}.

Before concluding this section we should remark that the set of CFC
equations we have discussed so far ignores radiation-reaction
terms~\cite{Faye2003, Oechslin07a}. These terms have been omitted mostly
to reduce the computational costs, because their contribution to the
spacetime dynamics is very small (see the \texttt{migration} test in
Sec.~\ref{sec:test_mt} or the \texttt{head-on} test in
Sec.~\ref{sec:head-on}), or because we perform the transfer of data
between the two codes tens of milliseconds after the merger, when
radiative GW contributions are already sufficiently small (see the
post-merger remnant test in Sec.~\ref{sec:test_hand_off}). However,
including these terms can be important in conditions of highly dynamical
spacetimes and would provide important information on the GW emission
from the scenarios simulated with \BHACns. Work is in progress to implement
these terms in the solution of the constraints sector and a discussion
will be presented elsewhere~\cite{Jiang2024:inprep}.

\section{Numerical Setup}
\label{sec:num_setup}
\subsection{Spacetime Solvers}
\label{sec:metric_init}

We briefly recall that, once the initial data has been
computed\footnote{In \FIL this is normally done using either the
open-source codes \texttt{FUKA}~\cite{Papenfort2021, Tootle2022},
\texttt{LORENE}~\cite{lorene} or the \texttt{COCAL}
code~\cite{Tsokaros2015, Most2019}.}, the spacetime solution in \FIL is
carried out in terms of the \textit{evolution sector} of the 3+1
decomposition of the Einstein equations~\cite{Alcubierre:2006,
  Rezzolla_book:2013} in conjunction with the
\texttt{EinsteinToolkit}~\cite{loeffler_2011_et,
  EinsteinToolkit_etal:2022_05}, exploiting the \texttt{Carpet}
box-in-box AMR driver in Cartesian coordinates~\cite{Schnetter:2006pg},
and the evolution code-suite developed in Frankfurt, which consists of
the \FIL code for the higher-order finite-difference solution of the
GRMHD equations and of the \texttt{Antelope} spacetime
solver~\cite{Most2019b} for the evolution of the constraint damping
formulation of the Z4 formulation of the Einstein
equations~\cite{Bernuzzi:2009ex, Alic:2011a}.

On the other hand, building on the xCFC scheme implemented in spherical,
cylindrical and Cartesian coordinates in the \texttt{Gmunu}
code~\cite{Cheong2020, Cheong2021}, \BHAC carries out the spacetime
solution in terms of the \textit{constraints sector} of the 3+1
decomposition of the Einstein equations~\cite{Alcubierre:2006,
  Rezzolla_book:2013} in the xCFC approximation. In essence, we solve the
set of elliptic xCFC equations using a cell-centered multigrid solver
(CCMG) \cite{Cheong2020, Cheong2021}, which is an efficient, low-memory
usage, cell-centered discretization for passing hydrodynamical variables
without any interpolation or extrapolation and can be coupled naturally
to the open-source multigrid library
\texttt{octree-mg}~\cite{Teunissen2019} employed by
\texttt{MPI-AMRVAC}~\cite{Porth2014} used by \BHACns. We recall that
multigrid approaches solve a set of elliptic partial differential
equations recursively, using coarser grids to efficiently compute the
low-frequency modes that are expensive to compute on high-resolution
grids (see, for instance, Ref.~\cite{Cheong2020} for more detailed
information). In addition, we employ the Schwarzschild solution for the
outer boundary conditions, using Eqs.~(77)--(82) in
Ref.~\cite{Cheong2021} and implementing Robin boundary conditions on the
cell-face for spherical polar coordinate and on the outermost cell-center
for cylindrical and Cartesian coordinates (see
Refs.~\cite{Bucciantini2011, Cheong2020, Cheong2021} for details).

As with any iterative scheme for the solution of an elliptic set of
partial differential equations, an accurate solution of metric variables
is determined when the infinity norm $L_{\infty}$ of the residual of the
metric equation, namely, the maximum absolute value of the residual of
the CFC equations, falls below a chosen tolerance. However, we need to
distinguish the tolerance employed for the solution of the initial
hypersurface (which may or may not coincide with the hypersurfaces
imported from \FILns) from the tolerance employed in the actual
evolution. More specifically, when importing data from \FIL at the
data-transfer stage and in order to minimize inconsistencies resulting
from different metric solvers or gauges used in the two codes, we set a
rather low tolerance, \ie $\varepsilon_{\mathrm{tol, in}} =
10^{-8}-10^{-10}$, depending on the type of initial data. During the
actual spacetime evolution, on the other hand, we strike a balance
between the computational costs and the accuracy of the solution of the
xCFC equations and increase the tolerance to $\varepsilon_{\mathrm{tol,
    ev}} = 10^{-6}$. As we will demonstrate later on, these choices
provide numerical solutions that are both accurate and computationally
efficient.

It is also important to remark that we modify the metric initialization
proposed by Ref.~\cite{Cordero2009} in which the values of $\psi$ are
iterated from an initial value of $\psi=\psi_0$ ($\psi_0 = 1$ initially
for most of the cases) and the conserved variables are obtained while
keeping fixed the initial primitive variables. However, this approach may
fail to converge to a proper value $\psi$ for extremely strong gravity
regions or for large gradient of the Lorentz factor. Another disadvantage
we have encountered is that this approach will lead to large deviations
between the ``handed-off'' data from \FIL and the newly converged
computed data by \BHACns. As a result, in our approach we first import
the gauge-independent quantities $\sqrt{\gamma/f} (D, S_j, \tau, DY_e,
B^j) = \psi^6 (D, S_j, \tau, DY_e, B^j)$, where $B^j$ is magnetic field
observed by an Eulerian observer, $Y_e$ is the electron fraction, $\gamma
:=\det{(\gamma_{ij})}$ and $f := \det{(f_{ij})}$. Next, we employ the
xCFC solver to compute all of the initial spacetime quantities. As we
will show in Sec.~\ref{sec:results}, this approach leads to initial data
whose evolution in \BHAC exhibits smaller deviations the corresponding
evolution from \FIL (see, \eg Fig.~\ref{fig:DD2BNS_1dslices}).

\subsection{Matter Solvers}
\label{sec:matter_solver}

As mentioned above, the solution of the GRMHD system of equations is
handled differently by the two codes in our hybrid approach, although
both of them follow high-resolution shock-capturing (HRSC)
methods~\cite{Toro99, Rezzolla_book:2013}. More specifically, the
Frankfurt/IllinoisGRMHD (\FILns) code is an extension of the publicly
available \texttt{IllinoisGRMHD} code~\cite{Etienne2015}, which utilizes
a fourth-order accurate conservative finite-difference
scheme~\citep{DelZanna2007}. On the other hand, \BHAC is a further
development of \texttt{BHAC} -- which itself was built as an extension of
the special-relativistic code \texttt{MPI-AMRVAC} -- to perform GRMHD
simulations of accretion flows in 1D, 2D and 3D on curved spacetimes
(both in general relativity and in other fixed
spacetimes~\citep{Mizuno2018, Olivares2020, Cruz2023}) using second-order
finite-volume methods and a variety of numerical methods described in
more detail in~\cite{Porth2017}. \texttt{BHAC} is publicly available and
has been employed in a number of applications to simulate accretion onto
supermassive black holes~\cite{Cruz2022}, compact stars~\cite{Das2022,
  Cikintoglu2022} and in dissipative hydrodynamics~\cite{Chabanov2021}.

Differently from \FILns, \BHAC exploit much of \texttt{MPI-AMRVAC}'s
infrastructure for parallelization and block-based automated AMR (see
Refs.~\cite{Porth2017, Olivares2019, Ripperda2019} for additional
details) employing a staggered-mesh upwind constrained transport schemes
to guarantee the divergence-free constraint of the magnetic
field~\cite{Londrillo2004, Olivares2019}. These methods represent an
improvement over the original constrained-transport
scheme~\cite{Evans1988} and aim at maintaining a divergence-free
condition with a precision comparable to floating-point operations,
ensuring that the sum of the magnetic fluxes through the surfaces
bounding a cell is zero up to machine precision. \FIL, on the other hand,
follows its predecessor \texttt{IllinoisGRMHD} code in computing the
evolution of the magnetic field via the use of a magnetic vector
potential, whose curl then provides the magnetic field. However,
differently from Ref.~\cite{Etienne2015}, \FIL implements the upwind
constraint-transport scheme suggested in Ref.~\citep{DelZanna2007}, in
which the staggered magnetic fields are reconstructed from two distinct
directions to the cell edges. This approach greatly minimizes diffusion
and cell-centred magnetic fields are always interpolated from the
staggered ones using fourth-order unlimited interpolation in the
direction in which the $i$-th component of the magnetic field is
continuous~\cite{Londrillo2004}. Overall, the two approaches implemented
in \BHAC and \FIL for handling the divergence-free constraint of the
magnetic field are overall equivalent both on uniformly spaced grids and
in grids with AMR levels. At the same time, a relevant difference between
the two codes in that \BHAC implements a new and robust primitive
recovery scheme with a tabulated EOS module, error-handling policy,
atmosphere treatment and the evolution equation of electronic lepton
number in order to account for an EOS that depends on temperature and
composition (see Sec.~\ref{sec:tabeos} for more details).

Before concluding this section, an important remark is worth making. A
fundamental aspect of our hybrid approach, and that leads to the single
most important advantage in terms of computational speed is that, unlike
typical free evolution schemes, the solution of the spacetime variables
in a constrained approach does not need to be performed on every
spacelike hypersurface on which the matter equations are solved. Indeed,
because the spacetime evolution takes place through the solution of a
system of elliptic equations, whose characteristic speed are not defined,
no stability constraint exists on the width of the temporal step. This is
to be contrasted with the solution of a system of hyperbolic equations --
such as those employed for the evolution of the Einstein equations in
\FIL and more generally for the matter sector in the two codes -- whose
characteristic speed are defined in terms of the light cone and where the
timestep is severely constrained by the Courant-Friedrichs-Lewy (CFL)
condition. As a result, the spacetime and matter solvers in \BHAC are
de-facto decoupled, and their update frequencies need not coincide.

This brings in two distinct advantages. The first one can be measured in
terms of the ``efficiency ratio'', $\xi_{\rm eff}:= N_{\rm spctm}/N_{\rm
  mattr}$, that is the ratio of spacetime timesteps over the matter
timesteps, which is necessarily $\xi_{\rm eff} = 1$ in typical
numerical-relativity evolution codes (\eg \FILns) that evolve the
Einstein-Euler equations as system of hyperbolic equations. On the other
hand, this ratio can be much smaller, \ie $\xi_{\rm eff} = 1-1/100$ in
codes evolving the Einstein-Euler equations as mixed system of
elliptic-hyperbolic equations (\eg \BHACns)\footnote{Note that in
constrained-evolution approaches, such as the one implemented in \BHACns,
the CFC equations are \textit{not} solved at each Runge-Kutta substep
when transitioning from time-level $n$ to time-level $n+1$. While this is
an approximation, a number of studies have shown that the differences in
the accuracy of the solution, when not updating the spacetime at each
substep, are negligible (see Refs.~\cite{Dimmelmeier02a, Bucciantini2011,
  Cheong2020}).}, with a gain in computational costs that is inversely
proportional to $\xi_{\rm eff}$. The second important advantage is that
during the matter-only evolution, the timestep, which is constrained by
the CFL factor and inversely proportional to the largest propagation
speed, is bounded by the speed of sound rather than the speed of light;
given that $c_s/c\simeq 0.1-0.3$, this fact alone provides an additional
and proportional reduction in computational costs.

Of course, these savings also come at the expense of some accuracy. For
instance, an excessively small $\xi_{\rm eff}$ ratio can lead to a slight
diffusion of matter out of the gravitational-potential well, which is not
updated frequently enough. A certain degree of experimentation is needed
to identify the optimal $\xi_{\rm eff}$ for a given scenario and we will
comment on this also later on. For the time being, we just mention that
in a $10\,\mathrm{ms}$ simulation of a 2D axisymmetric rapidly rotating
neutron star in spherical coordinates, a value of $\xi_{\rm eff} = 1/50$
is sufficient to capture most of the oscillation modes and even the
high-frequency ones~\cite{Cheong2020}.

\subsection{Spacetime and matter ``hand-off''}
\label{sec:test_hand_off}

Of course, an essential aspect of the hybrid approach discussed here is
represented by the so-called ``hand-off'' (HO), \ie the export of a
solution for the spacetime and fluid variables from \FIL to \BHAC and, in
principle, also in the other direction, although we will not discuss the
latter here.

In essence, the HO procedure in our approach can be summarized as follows
\begin{itemize}

\item at any specific time, \eg after the merger of a BNS system, we
  extract the primitive variables, along with the conformal factor, from
  the 3D \FIL data in order to to obtain the quantities $\sqrt{\gamma/f}
  (D, S_j, \tau, DY_e, B^j)$.

\item we transform the 3D data from the original Cartesian coordinates to
  a new coordinate system, which can be Cartesian, spherical polar or
  cylindrical depending on the system under investigation. In the case of
  a post-merger evolution we employ cylindrical coordinates as these are
  optimal for 2D axisymmetric evolutions (indeed, the cylindrical
  coordinates on a 2D constant-$\phi$ slice coincide with the Cartesian
  coordinates on a 2D constant-$x$ slice).

\item in the case of 2D \BHAC simulations, all quantities
  imported from \FIL are interpolated on a 3D cylindrical grid using
  simple linear interpolation (note that since \FIL is a
  finite-difference code, at this stage both metric and hydrodynamical
  numerical data can be interpreted as representing point-wise values
  of the respective fields at the grid coordinates). The data is
  subsequently averaged over the $\phi-$direction and vector and
  tensor variables are transformed to the coordinate system used for
  the evolution in \BHACns.

\item because the coordinates and grid-refinement structure used in
  \BHAC are obviously different from those in \FILns, it is possible
    that a high-resolution cell in \FIL is mapped to a low-resolution
    cell in \BHACns. To prevent this from happening, we always ensure
  that the resolution in the spatial region with $\rho > 10^{8} \,
  \mathrm{g~cm^{-3}}$ is either equal to or higher than that of the
  imported data.

\item we use the handed-off data to initialize the metric under the
  maximal-slicing gauge using the xCFC solver (see
  Sec.~\ref{sec:metric_init}), thus clearing any Hamiltonian and
  momentum-constraint violations that may have arisen due to the HO
  procedure.

\item we update the corresponding primitive variables under
  the CFC approximation.
\end{itemize}
Note that both \BHAC and \FIL adopt a conservative formulation of
  the GRMHD equations and hence the conserved variables are preserved at
  the level of machine precision. Violations of the conservation can only
  happen during the import of the data from \FIL to \BHAC and as a result
  of the coordinate remapping and interpolation. We have verified that
  the maximum relative differences in the import phase are below
  $0.4~\%$.

While much of the procedure described above applies also to the HO of 3D
\FIL data for a 3D \BHAC simulation, there are additional aspects --
besides the obvious skipping of the azimuthal averaging -- that need to
be taken into account when passing the data over to a 3D \BHAC grid to
ensure an optimal interpolation of all quantities and the preservation of
the divergence-free condition to machine precision. For compactness, and
because the HO presented here is from 3D \FIL to 2D \BHAC (see
Sec.~\ref{sec:longtermBNS}), we will omit such details and postpone their
discussion in a forthcoming companion paper~\cite{Jiang2024:inprep}.

It should also be noted that the conformal factor $\psi$ is a
gauge-dependent quantity and hence it exhibits differences between the
full numerical-relativity code \FIL and the CFC code
\BHACns\footnote{Note however that the quantities $\sqrt{\gamma/f} (D,
S_j, \tau, DY_e, B^j)$ are gauge-independent.}. Indeed the conformal
factor defined in a full numerical-relativity simulation assuming, say, a
$1+$~log-slicing reduces to the conformal factor used in the CFC scheme
with a maximal-slicing gauge only for systems for which the conformal
flatness represents a good approximation (\eg the initial data for
a BNS system). As we will discuss in the analysis of a BNS post-merger in
Sec.~\ref{sec:longtermBNS}, the comparison of the values of $\psi$
between the two approaches for the spacetime solution shows behaviours
that are very similar so that the use of the conformal factor represents
a simple and efficient way to compare spacetimes that approach conformal
flatness. However, a more rigorous and general approach could be offered
by the calculation and comparison of the values of the Cotton-York
tensor, which we will investigate in future analyses.

\subsection{Primitive-recovery scheme}
\label{sec:tabeos}

Obviously, the ability to handle realistic, temperature- and
composition-dependent EOSs is essential in order to achieve a realistic
description of the secular post-merger dynamics and hence arrive at
accurate predictions for multi-messenger astronomical observables from
merging BNSs, \eg GWs, gamma-ray burst signals, and kilonova
light-curves. Fully tabulated, nuclear-physics EOSs need to be employed
to this scope as they provide information on the pressure $p$ as a
function of the temperature $T$, the electron fraction $Y_e$, and
baryonic number (rest-mass) density ($\rho$) $n_b$, alongside with other
essential thermodynamic quantities, such as the baryon and lepton
chemical potentials, the speed of sound $c_s$, the specific entropy,
etc. Despite playing only a secondary role in the hierarchy of equations
to be solved, the use of these tabulated EOSs is far less trivial than it
may appear at first sight. The reason for this is the flux-conservative
formulation of the matter-evolution equations, which requires the
introduction of conserved variables that are distinct from the (physical)
primitive variables employed in the EOSs (see, \eg \cite{Banyuls97,
  Rezzolla_book:2013} for a discussion). The need to establish a bijective
mapping from one set to the other, and the nonlinear and non-analytic
nature of this mapping, makes the operation of primitive-recovery from
tabulated EOSs a major hurdle in modern codes, but also an important
aspect of code improvement and optimization.

In \FIL this problem is solved through the \texttt{Margherita}~
framework, a standalone modern \texttt{C++}~code that takes care of
reading and interpolating the EOS tables, as well as of the conservative
to primitive conversion. The latter is achieved in \texttt{Margherita} by
different procedures depending on the physical conditions at hand. For
unmagnetized fluids, \FIL employs the well known-and robust primitive
recovery scheme by Ref.~\cite{Galeazzi2013}. If magnetic fields are
non-negligible in the fluid, the inversion is performed with the
one-dimensional algorithm by \citet{Palenzuela2015}. In case of failure
of any of the primary methods, the entropy is used instead of the
temperature to correctly recover the primitive variables from the
conserved ones. In \BHACns, on the other hand, the inclusion of
temperature-dependent EOSs has been accomplished only recently, since
\texttt{BHAC} only allowed for the use of analytic EOSs (\ie ideal-fluid,
Synge gas, isentropic flow~\cite{Porth2017}). Hence, considerable work
has been invested in extending the capabilities of \BHAC to handle
generic EOSs and, more importantly, to obtain a framework that provides a
robust primitive-variable recovery with finite-temperature tabulated
EOSs. Currently, \BHAC can support tabulated EOS in either the format of
the \texttt{StellarCollapse}~\cite{Stellarcollapse} or in that of the
\texttt{CompOSE} repository~\cite{Typel2015}.

In essence, all the thermodynamic quantities $Q$ are assumed to be
calculated under local thermodynamical equilibrium and are expressed as
functions of $\rho, T, Y_e$ in CGS units (the temperature is normally
expressed in $\mathrm{MeV}$), although they are transformed to code units
for convenience. Inevitably, these tables may contain unphysical values
and thus ensuring the validity of all the thermodynamical quantities is
crucial both to achieve stable evolution and for accurate estimates of
neutrino opacities. To address this issue, we have implemented checkers
for every table in order to identify and handle unphysical values
appropriately. For instance, we ensure that the sound speed satisfies the
obvious condition $0 \leq c_s^2 \leq 1$, but we also determine for each
tabulated quantity the corresponding minimum and maximum bounds, \ie
$\rho_{\mathrm{min/max}}$, $\epsilon_{\mathrm{min/max}}$,
$T_{\mathrm{min/max}}$, $Y_{e,\mathrm{min/max}}$, and
$h_{\mathrm{min/max}}$. As we comment below, these bounds will be useful
for a robust treatment of the atmosphere and for an accurate
primitive-recovery scheme.

In this context, various algorithms have been developed over the years to
ensure an accurate, efficient, and stable primitive recovery, aiming at
minimizing error accumulation during the matter evolution. Comparisons of
different algorithms have been studied in Refs.~\cite{Siegel2018a,
  Espino2022} with specific focus on their accuracy and robustness. Among
the numerous primitive-recovery algorithms, the one developed by
Kastaun~\cite{Kastaun2021} has been extensively investigated and
demonstrated several advantages in GRMHD simulations with analytical
EOSs. More specifically, it employs a smooth, one-dimensional,
continuous, and well-developed master function that guarantees that a
root is found within a given interval and the uniqueness of the solution
is ensured even for unphysical values of the conserved variables. This
procedure does not require derivatives of the EOS or an initial guess,
thereby making it particularly efficient and robust, showing high
accuracy in regimes with high Lorentz factors and strong magnetic fields,
as well as low-density environments where fluid-to-magnetic pressure
ratios can reach values as low as $10^{-4}$ (see ~\cite{Kastaun2021} for
more details).

This approach has been successfully implemented in GRMHD codes such as
\texttt{Gmunu}~\cite{Cheong2021, Cheong2023, Ng2023a}, \texttt{ReprimAnd}
within the \texttt{Einstein Toolkit}~\cite{Loffler:2011ay}, although
  only for analytic EOSs, and more recently within the
  \texttt{GR-Athena++} code~\cite{Cook2023} to include tabulated EOSs,
  where the temperature and electron fraction serves as additional
  independent variables. In what follows, we illustrate in detail and in
a sequential manner the adaptations of Kastaun's algorithm that are
needed for its application in simulations with tabulated
EOSs. Furthermore, we present for the first time a systematic
  assessment of its robustness and efficiency with fully tabulated EOSs.

\textit{(i)} We first calculate the electron fraction $Y_e$ using the two
conserved quantities $D:=\rho W$, which is the conserved rest-mass
density, and $DY_e$. In other words, we compute $Y_e=DY_e/D$ and consider
it within the specified bounds given by $Y_{e,\mathrm{min}} \leq Y_e \leq
Y_{e,\mathrm{max}}$. If $D$ falls below a defined threshold value, \ie $D
< D_{\mathrm{thr}} = \rho_{\mathrm{thr}}$, where $\rho_{\mathrm{thr}}$
denotes the atmospheric threshold of rest-mass density (see
Sec.~\ref{sec:atmo}), we consider the corresponding numerical cell as
part of the atmosphere and skip the entire primitive-recovery process to
minimize computational costs.

\textit{(ii)} We next introduce the rescaled conserved variables defined
as
\begin{equation}
q := \frac{\tau}{D}\,, \qquad r_i := \frac{S_i}{D}\,, \qquad
\mathcal{B}^i := \frac{B^i}{\sqrt{D}}\,,
\end{equation}
noting that in the ideal-MHD limit, the magnetic field observed by an
Eulerian observer $B^i$ is either an evolved variable or can be
reconstructed from the evolved variables without requiring knowledge of
the fluid-related primitive variables. We further decompose the rescaled
momentum into the components parallel and perpendicular to the magnetic
field, namely
\begin{equation}
  r_{\|}^i := \frac{\mathcal{B}^j r_j}{\mathcal{B}^2} \mathcal{B}^i\,,
  \qquad r_{\perp}^i := r^i - r_{\|}^i\,.
\end{equation}

\textit{(iii)} We setup an auxiliary function defined as 
\begin{equation}
  f_a(\mu) := \mu \sqrt{h_{\mathrm{min}}^2 +
    \bar{r}^2(\mu)} - 1\,,
  \label{eq:aux_func}
\end{equation}
where $h_{\mathrm{min}}$ is the minimum value of specific enthalpy as
derived from the tabulated EOS (\cf Sec.~\ref{sec:tabeos}). The
quantities $\bar{r}^2(\mu)$, $\chi(\mu)$, and $\mu$ are instead defined
as
\begin{align}
  \bar{r}^2(\mu) &:= r^2 \chi^2(\mu) + \mu \chi(\mu)(1+\chi(\mu))
  \left(r^j \mathcal{B}_j\right)^2\,, \\
\chi(\mu) &:= \frac{1}{1+\mu \mathcal{B}^2}\,, \\
\mu &:= \frac{1}{hW}\,,
\end{align}
where $r^2 = r^i r_i$ and $\mathcal{B}^2 = \mathcal{B}^i\mathcal{B}_i$,
and $\mu$ is restricted to the range $0<\mu \leq 1 / h_{\mathrm{min}}$.
To find the root $\mu_+$ of $f_a(\mu)$ we employ a Newton-Raphson
root-finder method within the interval $\mu \in (0,
1/h_{\mathrm{min}}]$. Since $f_a(\mu)$ is a smooth function and does not
  require calls to the tabulated EOS, its derivative can be determined
  analytically. In this way, we can efficiently obtain an useful initial
  bracketing of the root of the master function [see
    Eq.~(\ref{eq:Kmaster})] in the interval $(0, \mu_+]$ and ensure that
    the condition $v \leq v_0<1$ is satisfied, where
\begin{equation}
  v:= \mu \bar{r}, \qquad
  v_0:= \frac{r^2}{h_\mathrm{min}^2 + r^2}\,. \label{eq:v0}
\end{equation}

\textit{(iv)} Next, we solve the one-dimensional master function $f(\mu)$ 
\begin{equation}
  \label{eq:Kmaster}
  f(\mu):=\mu-\frac{1}{\mathrm{max}(\nu_A, \nu_B) + \mu \bar{r}^2(\mu)}\,,
\end{equation}
in the bracketed interval $\mu \in (0, \mu_{+}]$ using Brent's
  method~\cite{Brent2002}. The master function $f(\mu)$ depends on the
  variables listed below, which are calculated in the following order
\begin{align}
  &\bar{q}(\mu)=q-\frac{1}{2} \mathcal{B}^2-\frac{1}{2} \mu^2 \chi^2(\mu)\left(\mathcal{B}^2 r_{\perp}^2\right)\label{eq:q_bar} \,, \\
  &\hat{v}^2(\mu)=\min \left(\mu^2 \bar{r}^2(\mu), v_0^2\right)\label{eq:v2_hat} \,, \\
  &\hat{W}(\mu)=\frac{1}{\sqrt{1-\hat{v}^2(\mu)}}\label{eq:W_hat} \,, \\
  &\hat{\rho}_0(\mu)=\frac{D}{\hat{W}(\mu)}\label{eq:rho_hat} \,, \\
  &\hat{\rho}(\mu)= \mathrm{max}[\rho_{\mathrm{min}},\mathrm{min}(\rho_{\mathrm{max}},\hat{\rho}_0)]  \label{eq:boundofrho} \,, \\
  &\hat{\epsilon}_0(\mu)=\hat{W}(\mu)\left(\bar{q}(\mu)-\mu \bar{r}^2(\mu)\right)+\hat{v}^2(\mu) \frac{\hat{W}^2(\mu)}{1+\hat{W}(\mu)}\label{eq:eps_hat} \,, \\
  &\hat{\epsilon}(\mu) = \mathrm{max}[\hat{\epsilon}_{\mathrm{low}}(\hat{\rho},Y_e),\mathrm{min}(\hat{\epsilon}_{\mathrm{high}}(\hat{\rho},Y_e),\hat{\epsilon}_0)] \label{eq:boundofeps}\,, \\
  &\hat{p}=p(\hat{\rho}, \hat{T}(\hat{\rho},\hat{\epsilon},Y_e), Y_e)\label{eq:p_hat} \,, \\
  &\hat{a}(\mu)=\frac{\hat{p}}{\hat{\rho}(\mu)(1+\hat{\epsilon}(\mu))}\label{eq:a_hat} \,, \\
  &\nu_A(\mu)=(1+\hat{a}(\mu)) \frac{1+\hat{\epsilon}(\mu)}{\hat{W}(\mu)}\label{eq:v_A} \,, \\
  &\nu_B(\mu)=(1+\hat{a}(\mu))\left(1+\bar{q}(\mu)-\mu \bar{r}^2(\mu)\right)\label{eq:v_B} \,. 
\end{align}

In Eq.~(\ref{eq:boundofrho}) we ensure that $\hat{\rho}$ remains within
the bounds of the table during each iteration, while we define
$\hat{\epsilon}_{\mathrm{low/high}} := \hat{\epsilon}(\hat{\rho},
T_{\mathrm{min/max}},Y_e)$\footnote{Note that the adjectives ``low/high''
should not be confused with the adjectives ``min/max''. The latter refer
to the ranges in the table, while the former refer to the minimum and
maximum values within the iteration.}  to guarantee that $\hat{\epsilon}$
is properly bracketed for the root-finding inversion of
$\hat{\epsilon}(\hat{\rho},\hat{T},Y_e)$ to $\hat{T}(\hat{\rho},
\hat{\epsilon}, Y_e)$, which is needed in Eq.~(\ref{eq:p_hat}). A value
for $\hat{T}$ is thus found by solving for the root of the function
\begin{equation}
  \label{eq:epstoT_aster}
  f(T_i)=1-\epsilon_i(\hat{\rho},T_i,Y_e)/\hat{\epsilon} ,
\end{equation}
within the interval $T_i \in [T_{\mathrm{min}}, T_{\mathrm{max}}]$ using
Brent's method. We note that oscillating unphysical values of
$\hat{\epsilon}$ within the root-finding iteration of
Eq.~(\ref{eq:Kmaster}) can at times prevent the determination of a root.
When the number of iterations exceeds a certain threshold (we set this
  to be $50$), we return $\hat{\epsilon} = \epsilon_{\mathrm{low}}
(\hat{\rho}, T_{\mathrm{min}}, Y_{e})$ and $\hat{T}=T_{\mathrm{min}}$ if
$\hat{\epsilon} \leq \epsilon_{\mathrm{low}} (\hat{\rho},
T_{\mathrm{min}}, Y_{e})$ or $\hat{\epsilon} =
\epsilon_{\mathrm{high}}(\hat{\rho}, T_{\mathrm{max}}, Y_{e})$ and
$\hat{T}=T_{\mathrm{max}}$ if $\hat{\epsilon} \geq
\epsilon_{\mathrm{high}}(\hat{\rho}, T_{\mathrm{min}}, Y_{e})$.

\textit{(v)} The subsequent step involves using the converged root $\mu$
obtained from Eq.~(\ref{eq:Kmaster}), with a specified tolerance, to
determine the primitive variables as listed in the previous step. For the
calculation of the velocity $\hat{v}^i$, in particular, we use
\begin{equation}
\hat{v}^i(\mu) = \mu \chi(\mu)\left(r^i+\mu\left(r^j \mathcal{B}_j\right)
\mathcal{B}^i\right)\,.
\end{equation}
Once this stage is reached, we check for cells falling into the
atmosphere and to them apply the error-handling policy presented below in
Sec.~\ref{sec:atmo}.

\textit{(vi)} Finally, with the updated values of $\rho$, $T$ and $Y_e$,
we can obtain $c_s^2$, $p$, as well as any other required thermodynamic
quantity by a EOS call without invoking one more time of inversion of
$\epsilon$ to $T$. At the end, we recalculate the corresponding conserved
variables to ensure they are consistent with the updated primitive
variables. This step is important considering that the EOS routine, the
atmospheric treatment, or safety checks may have modified the primitive
variables.

We note that when following the steps \textit{(i)}--\textit{(vi)} in the
algorithm presented above, no restrictions are made on negative values of
the specific internal energy $\epsilon$ or on values of the specific
enthalpy being $h < 1$, which are possible when the (negative) nuclear
binding energy exceeds the thermal or excitation energy. These values,
however, could pose a problem during the inversion between $\epsilon$ and
$T$ at each intermediate step. More specifically, when $\epsilon$ does
not increase monotonically with $T$ -- which can be the case in the low
$\epsilon$ range of tabulated EOSs -- incorrect values of $T$ can be
obtained. Our approach to counter these cases is to input an initial
guess for temperature which is obtained from the last result 
in the root-finding method and to update this guess
throughout the primitive-recovery procedure. For achieving full
consistency between \BHAC and \FIL in the tests to be presented in the
following sections, the conservative to primitive conversion procedure
outlined above was also implemented in \texttt{Margherita}.

\begin{figure*}
  \includegraphics[width=1.0\textwidth]{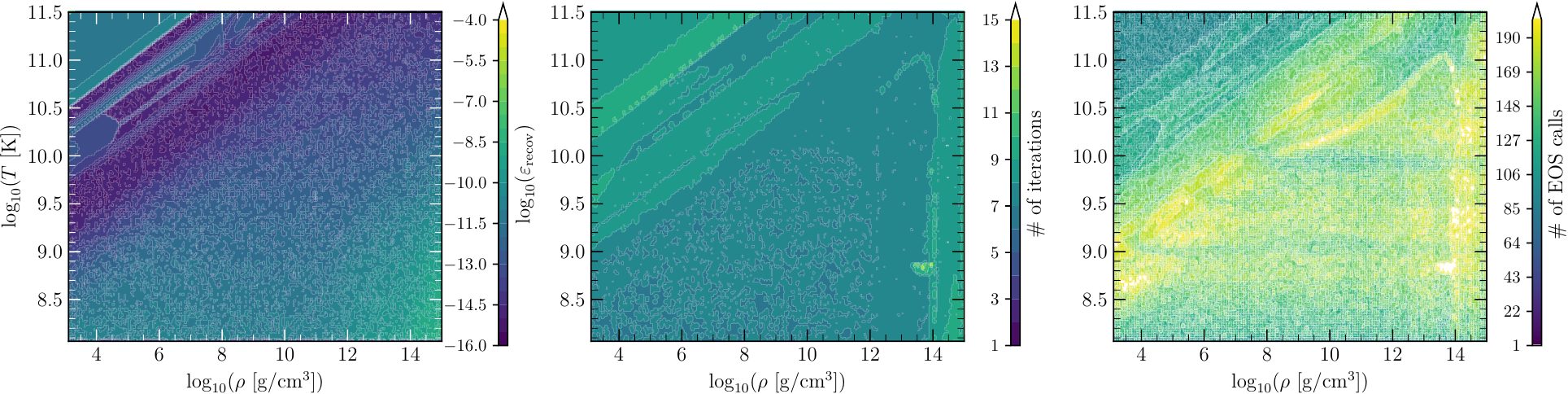}
  \caption{Average relative error $\varepsilon_{\rm recov}$ (left panel),
    number of iterations (middle panel), and EOS calls (right panel)
    required to reach the desired tolerance in recovering the primitive
    variables from the conserved ones. The example shown employs the
    LS220 EOS ~\cite{Lattimer91} with parameter values $W = 2$ and $Y_e =
    0.1$. The magnetic field is nonzero and set so that the pressure
    ratio is $b^2/2p = 10^{-3}$, with $b^2:=b_ib^i$ being the strength of
    the magnetic field in the fluid frame.}
  \label{fig:Kc2p_performance}
\end{figure*}

\subsection{Performance of the primitive-recovery scheme}

We here evaluate the performance of our primitive-recovery scheme
presented in the previous section in terms of accuracy and efficiency,
and compare it to other schemes used in GRMHD simulations and that are
either referred to as 1D~\cite{Neilsen2014, Newman2014,
  Palenzuela2015}, 2D~\cite{Noble2006, Anton05}, or 3D~\cite{Cerda2008},
depending on the dimensionality of the master root-finding
function. Furthermore, to ensure a fair comparison with previous
primitive-recovery schemes that use tabulated EOS, we adopt two of the
tests mentioned in~\cite{Siegel2018a} and follow the same criteria
outlined there, which include considerations of speed, accuracy, and
robustness (see Sec.~4.1 of~\cite{Siegel2018a} for additional
information). In both tests considered here -- and for consistency with
other previously published results -- we have used the LS220 tabulated
EOS~\cite{Lattimer91}.

In the first test, we use ranges of $\rho$ and $T$ that cover the valid
regions of the EOS table. More specifically, we select primitive
variables with the following values: a Lorentz factor of $W = 2$, a ratio
of magnetic-to-fluid pressure of $b^2/(2p) = 10^{-3}$, and an electron
fraction of $Y_e = 0.1$. Figure~\ref{fig:Kc2p_performance} presents the
average relative error as a function of the number of iterations and EOS
calls required for convergence. More precisely, we compute the average
relative error as~\cite{Siegel2018a}
\begin{equation}
  \varepsilon_{\rm recov} := \frac{1}{5} \sum^{5}_{i=1} \left| 1 -
  \frac{\boldsymbol{P}_{i,\mathrm{recov}}}{\boldsymbol{P}_{i,
      \mathrm{orig}}} \right|\,,
\end{equation}
where $\boldsymbol{P}_{i,\mathrm{recov}}$ refers to the five recovered
primitive variables $(\rho, v^i, \epsilon, B^i, Y_e)$, while
$\boldsymbol{P}_{i,\mathrm{orig}}$ indicates the original values of the
primitives. Furthermore, we stop the iterations when a residual error of
$\leq 5 \times 10^{-9}$ is obtained for the maximum relative error in the
iteration variables through our primitive-recovery scheme. Overall,
Fig.~\ref{fig:Kc2p_performance} illustrates that across the entire
parameter space encompassing $\rho$ and $T$, the average relative error
$\varepsilon_{\rm recov}$, the average number of iterations, and the
average number of EOS calls are found to be $7.84 \times 10^{-12}$,
$7.989$, and $135.8$, respectively.

With these results, and before entering in the details of the comparison,
it is worth noting that multi-dimensional recovery schemes tend to
require fewer EOS calls (about 3-8 times less) compared to 1D schemes
(this was discussed also in~\cite{Siegel2018a}), but also a similar
number of iterations (5-9), in order to reach converge. The high number
of EOS calls in the effective 1D schemes is primarily due to the
additional inversion steps caused by the use of the EOS table in terms of
$T$ instead of $\epsilon$. Therefore, the number of EOS calls, and the
associated computationally expensive interpolations, the table look-ups,
and the root-finding procedures for the inversion of $\epsilon$ to $T$,
can be taken as a direct proxy of the numerical costs.

A similar behaviour, \ie few iterations, many EOS calls, is found also
with our recovery scheme, which is effectively a 1D scheme with an
additional inversion step from $\epsilon$ to $T$ using the table.
However, when comparing our recovery scheme with the other schemes
discussed in Ref.~\cite{Siegel2018a}, we have found a clear improvement
in terms of efficiency, as our approach requires significantly fewer EOS
calls. At the same time, although our scheme requires a number of
iterations that is similar to that reported in
Refs.~\cite{Palenzuela2015,Newman2014}, the mean number of EOS calls is
$135.8$, which is to be compared respectively with $836$ for
Ref.~\cite{Palenzuela2015} and 331 for Ref.~\cite{Newman2014}. In
addition, our scheme exhibits a lower average relative error when
compared to all other schemes, in particular within the regime relevant
to realistic astrophysical problems, such as for rest-mass densities in
the range $\rho \in [10^{8},10^{14}]\,\mathrm{g~cm^{-3}}$ and for the
entire range of temperatures $T_{\rm min/max}$ of typical tabulated
EOSs. More precisely, the schemes in Refs.~\cite{Cerda2008}
and~\cite{Newman2014} yield the lowest average relative error among all
the schemes, with values of $1.3 \times 10^{-13}$ and $6.1 \times
10^{-13}$, respectively. However, the accuracy in these schemes is not
homogeneous and much higher in the upper left corner of the parameter
space, while the relative error increases significantly for rest-mass
densities typical of neutron stars. On the other hand, our scheme covers
with high accuracy the entire parameter space with fewer than 12
iterations, except for a few points that require a larger number of
iterations for convergence. Finally, no failures are found in contrast to
what experienced with other schemes.

\begin{figure}
\center
\includegraphics[width=0.49\textwidth]{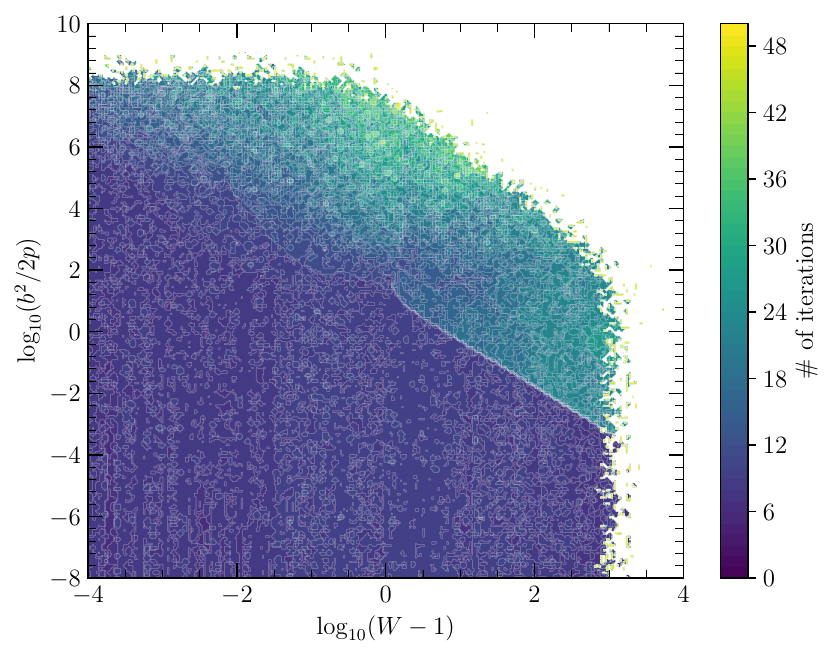}
\caption{Number of iterations required to reach the desired tolerance in
  recovering the primitive variables from the conserved ones when using
  the LS220 EOS with parameters values $\rho =
  10^{11}\,\mathrm{g~cm^{-3}}$, $T = 5\,\mathrm{MeV} = 5.8 \times
  10^{10}\,\mathrm{K}$, and $Y_e = 0.1$.}
\label{fig:Kc2p_performance_fixedTrho}
\end{figure}

To further establish the robustness of our primitive-recovery scheme, we
again follow~\cite{Siegel2018a} and conduct a second test that iterates
over the parameter space of the Lorentz factor $W$ and the ratio of
magnetic-to-fluid pressure $b^2/2p$, while maintaining fixed the values
of $\rho = 10^{11}\,\mathrm{g~cm^{-3}}$, $T = 5\,\mathrm{MeV}$, and
$Y_{e} = 0.1$. Figure~\ref{fig:Kc2p_performance_fixedTrho} illustrates
the number of iterations required for convergence in this second test and
shows that the maximum number of iterations in this parameter space is
$51$; the white spaces indicate areas where the recovery process failed
or the desired tolerance could not be achieved with the corresponding
parameter sets. In full analogy with the performance of other schemes
considered in~\cite{Siegel2018a}, also our recovery scheme fails when the
fluid becomes ultra-magnetized, \ie when $b^2/2p \gtrsim 10^{8}$, or when
the flow is ultra relativistic, \ie $W \gtrsim 10^3$. However, our
recovery scheme performs better than all other schemes, which either
failed during recovery or required more than $25$ iterations to achieve a
recovery when $W > 10 - 100$.

In terms of robustness, our scheme exhibits a similar performance to that
reported in Ref.~\cite{Newman2014}, successfully recovering the primitive
variables at $W = 1000$, except when $b^2/2p > 10^{-2}$. In such cases,
the efficiency slightly degrades, and the number of iterations increases
to $20-25$ for $W > 10$. However, since it does not require initial
guesses, or thermodynamic derivatives, or an initial bracket for the root
$\mu_{+}$ in Eq.~(\ref{eq:aux_func}), and guarantees the existence and
uniqueness of the root~\cite{Kastaun2021}, our recovery scheme can be
employed reliably over a larger parameter space when compared to most
other recovery schemes.

In summary, on the basis of the battery of sets performed, we conclude
that the primitive-recovery scheme presented in Sec.~\ref{sec:tabeos}
provides higher robustness and accuracy when compared to other 1D and
multi-dimensional schemes reported in Ref.~\cite{Siegel2018a} and
requires the smallest number of EOS calls among all the effective 1D
schemes. As a consequence of its robustness, no fail-safe strategy in the
case of failed primitive recovery is needed in \BHACns. We also note
  that Ref.~\cite{Cook2023} has very recently presented a
  primitive-recovery algorithm that is rather similar to the one
  presented here, with the main difference that the total energy density
  $e$ is used instead of the specific internal energy $\epsilon$ for the
  iteration.

\subsection{Atmosphere treatment and error-handling policy}
\label{sec:atmo}

In analogy with other codes, the ``atmosphere'' -- \ie the spatial region
of very low rest-mass density needed to avoid the failure of the solution
of the GRMHD equations -- is characterized by two basic parameters: the
rest-mass density threshold, denoted as $\rho_{\mathrm{thr}}$, and the
ratio between the density threshold and the atmosphere density $\xi :=
\rho_{\mathrm{atm}} / \rho_{\mathrm{thr}}$. In all simulations conducted
with \BHACns, these parameters are typically set to $\rho_{\mathrm{thr}}
= 10^{-14}$ and $\xi = 0.9$, respectively. Furthermore, all cells with
$\rho < \rho_{\mathrm{thr}}$ are identified as atmosphere and the
corresponding values of the primitives variables are set as follows:
\begin{align}
  &\rho = \rho_{\mathrm{atm}}\,,& &W = 1\,,&  &v^{i} = 0\,, 
  \nonumber \\
  &\epsilon = \epsilon_{\mathrm{atm}}\,,& &Y_e = Y_{e,\mathrm{atm}}\,, & 
  &p = p_{\mathrm{atm}}\,, 
  \nonumber \\
  & T = T_{\mathrm{atm}}\,, & &c_s^2 = c^2_{s,\mathrm{atm}}\,, &
  & B= B_{\mathrm{atm}}
  \,. \nonumber
\end{align}
Since a cell falling in the atmosphere is set to have zero velocity,
  the induction equation in the ideal-MHD limit prevents the evolution of
  the corresponding magnetic field. Hence, apart from changes due to the
  shift, $B_{\mathrm{atm}}$ can change only if the cell is pushed out of
  the atmosphere conditions via a nonzero velocity, or an increase in the
  rest-mass density above $\rho_{\rm atm}$.

When considering the atmosphere treatment in \BHACns, we need to
distinguish the situation in which an analytical EOS is used from that in
which the EOS is tabulated. In the former case, the polytropic EOS is
applied to describe the properties of the atmosphere, so that the
pressure is given by
\begin{equation}
p_{\mathrm{atm}}=K \rho_{\mathrm{atm}}^\Gamma,
\end{equation}
with $K = 100$ and $\Gamma = 2$ for the polytropic constant and
polytropic index, respectively. On the other hand, the specific internal
energy and the square of the sound speed in the atmosphere can be
obtained analytically as (see, \eg~\cite{Rezzolla_book:2013})
\begin{eqnarray}
  \epsilon_{\mathrm{atm}}&=&\frac{K
    \rho_{\mathrm{atm}}^{\Gamma-1}}{\Gamma-1}\,,
\nonumber\\
c^2_{s,\mathrm{atm}} &=&
\frac{p_{\mathrm{atm}} \Gamma (\Gamma - 1)}{\rho_{\mathrm{atm}} (\Gamma -
  1)+ p_{\mathrm{atm}} \Gamma}\,.
\nonumber
\end{eqnarray}

A different approach needs to be adopted when utilizing a tabulated EOS,
in which case the neutrinoless $\beta$ equilibrium condition is
employed. More specifically, at the beginning of the simulation, after
setting $\rho_{\mathrm{atm}}$ and $T_{\mathrm{atm}} = T_{\mathrm{min}}$,
a root-finding process is performed to determine the value of the
electron fraction $Y^{\beta}_{e}$ that satisfies the neutrinoless $\beta$
equilibrium condition
\begin{equation}
0 = \mu_e(Y_{e,i}) + \mu_p(Y_{e,i}) - \mu_n(Y_{e,i}) \,,
\end{equation}
where we employ Brent's method within the interval $Y_{e,i} \in
[Y_{e,\mathrm{min}}, Y_{e,\mathrm{max}}]$ and with $\mu_{e/p/n}$
representing the chemical potentials accounting for the rest-mass of
electrons, protons, and neutrons, respectively. Once $Y^{\beta}_{e}$ is
determined, it is adopted as the electron fraction value for the
atmosphere, $Y_{e,\mathrm{atm}}$. The remaining atmospheric quantities,
namely $\epsilon_{\mathrm{atm}}$, $p_{\mathrm{atm}}$, and
$c^2_{s,\mathrm{atm}}$, can be obtained through the EOS using
$\rho_{\mathrm{atm}}$, $T_\mathrm{atm}$, and $Y_{e,\mathrm{atm}}$ for
neutrinoless $\beta$-equilibrium matter.

\subsection{Error-handling policy}
\label{sec:err_policy}

It is not uncommon in modern relativistic GRMHD codes that physical
conditions of low rest-mass density and high magnetization may lead to
the generation of unphysical values of the matter quantities, especially
in regions that are treated as atmosphere, or where round-off errors may
develop, \eg near the surface of compact objects or in ultra-relativistic
flows. In order to ensure stable long-term simulations and maintain
accurate evolutions, error-handling procedures play a crucial role in
determining the criteria to be followed first to flag a problematic cell
and second to correct its physical representation. In our strategy for
treating problematic fluid cells, we incorporate some of the
prescriptions described in Ref.~\cite{Galeazzi2013} and adopt the
following list of error-handling policies:

\begin{enumerate}
  
\item if $\rho > \rho_{\mathrm{max}}$, then mark as a fatal error.

\item if $\epsilon < \epsilon_{\mathrm{min}}$, then set $\epsilon =
  \epsilon_{\mathrm{min}}$.

\item if $\epsilon > \epsilon_{\mathrm{max}}$, then mark as a fatal error.

\item if $T < T_{\mathrm{min}}$, then set $T = T_{\mathrm{min}}$.

\item if $T > T_{\mathrm{max}}$, then mark as a fatal error.

\item if $Y_e < Y_{e,\mathrm{min}}$, then set $Y_e = Y_{e,\mathrm{min}}$.

\item if $Y_e > Y_{e,\mathrm{max}}$, then set $Y_e = Y_{e,\mathrm{max}}$.

\item if $\epsilon \leq \epsilon_\mathrm{low}
  (\rho,T_{\mathrm{min}},Y_e)$, then within the inversion of $\epsilon$
  to $T$ set $\epsilon = \epsilon_\mathrm{low}$ and $T = T_\mathrm{min}$.

\item if $\epsilon \geq \epsilon_\mathrm{high} (\rho, T_{\mathrm{max}},
  Y_e)$, then within the inversion step of $\epsilon$ to $T$ set
  $\epsilon = \epsilon_\mathrm{high}$ and $T = T_\mathrm{max}$.

\item if $W > W_{\mathrm{max}}$, then, only in relatively low
  rest-mass density regions (\ie for $\rho \leq 10^{11}\,{\rm
      g~cm^{-3}}$), we limit $W = W_{\mathrm{max}}$ and set $v =
  v_{\mathrm{max}} := \sqrt{1 - 1/W^2_{\mathrm{max}}}$. The conserved
  density $D$ is kept fixed and we calculate $\rho = D/W_{\mathrm{max}}$.
\end{enumerate}

Finally, while these policies are fully generic, they are systematically
applied in the following five different scenarios:
\begin{enumerate}
\item after importing initial data;
\item before the inversion from $\epsilon$ to $T$;
\item after primitive recovery; 
\item after reconstruction from the left and right-hand sides;
\item after restriction or prolongation steps following the mesh refinement.
\end{enumerate}

\section{Results}
\label{sec:results}

We next present a series of numerical tests aimed at assessing not only
the accuracy and stability of \BHACns, but also its ability to import a
timeslice from a fully numerical-relativity simulation provided by \FIL
and evolve it stably for timescales up to one second. These
representative tests simulate various astrophysical systems with
increasing degree of realism and complexity, hence, starting from
oscillating nonrotating stars, to go over to rapidly rotating stars,
magnetized and differentially rotating stars, the head-on collision of
two neutron stars, and to conclude with the long-term BNS post-merger
remnants. The tests span across different spatial dimensions, ranging
from 1D to 2D and 3D, employing either spherical, cylindrical, or
Cartesian coordinates, and different types of EOSs, from analytical to
tabulated, as well as with varying spacetime conditions, including both
fixed and dynamical spacetimes. Table~\ref{tab:ID} provides a summary of
the parameters, dimensions, coordinate systems, and grid information for
each set of initial data used in the various tests.

\subsection{Tests setup and initial conditions}
\label{sec:setup}

In all simulations performed using \FILns, the time integration of the
full system of Einstein-Euler equations is performed using a Method of
Lines (MOL)~\cite{Rezzolla_book:2013}, with a third-order Runge-Kutta
method and a fixed CFL factor of $\mathcal{C}_{_{\rm CFL}}=0.2$ [see
    Eq. (8.41) of Ref.~\cite{Rezzolla_book:2013} for a definition]. The
GRMHD equations are solved with a two-wave Harten-Lax-van Leer-Einfeldt
(HLLE) Riemann solver~\cite{Harten83, Einfeldt88}, and a WENO-Z (Weighted
Essentially Non-Oscillatory with Z characteristic)
reconstruction~\citep{Borges2008} coupled to an HLLE Riemann
solver~\cite{Harten83} (see Refs.~\cite{Most2019b, Most2020e,
  Chabanov2022} for additional details). On the other hand, all the
simulations performed with \BHAC employ the Harten-Lax-van Leer (HLL)
Riemann solver~\cite{Harten83}, a Piecewise Parabolic Method
(PPM)~\cite{Colella84}, and a third-order Runge-Kutta (RK3) time
integrator. A second-order Lagrange interpolation is used for the metric
interpolation and the divergence-free constraint is enforced by using
upwind constrained transport~\cite{Olivares2019}. Information on the
coordinates used, the computational domain, the resolution on the
coarsest level, and the number of refinement levels employed is presented
in Table~\ref{tab:ID} for each simulation. For the adaptivity in the
mesh-refinement process, we employ a L{\"o}hner error
estimator~\cite{Loehner87} based on the values of the rest-mass density.

Special and different care needs to be paid depending on the type of
coordinates used for the simulations in \BHACns. In particular, in the
case of cylindrical coordinates, the highest refinement level is set to
be within a spherical region of radius $R < R_{\rm{in}} = 30\,M_{\odot}
\simeq 44.3\,\mathrm{km}$ and this is always sufficient to resolve the
high-density region of the studied systems. In the case of spherical
coordinates, however, a different strategy is necessary.

This is because the very small spatial size $\Delta x$ of the
  computational cells near the center of the coordinate system sets
  challenging constraints on the size of a CFL-stable timestep (we recall
  that $\Delta t \propto \mathcal{C}_{_{\rm CFL}} \Delta x$). For this
reason, in the case of spherical coordinates we employ a single grid
block covering an inner spherical region of radius $R_{\rm{core}} <
1-2\,M_{\odot} \simeq 1.48-2.95\,\mathrm{km}$, which is not the first
(highest) refinement level but the second one; this allows us to have
good resolution near the coordinate center but not to be penalized by an
excessively small timestep. The first refinement level is instead set
within the spherical shell with $R_{\rm{core}} < R < R_{\rm{in}}=
30\,M_{\odot} \simeq 44.3\,\mathrm{km}$, while the third (lowest)
refinement levels covers the region with $R_{\rm{in}} < R < R_{\rm out}$,
where $R_{\rm out}$ is the maximum value of the radial coordinate (see
Tab.~\ref{tab:ID} where $R_{\rm out}=x^r_{\rm max}$), and thus contains
the outer boundaries. This setup, while not fully exploiting the
adaptive-mesh capabilities of \BHACns, results in a reasonable timestep
constraint. Additionally, one of our tests (\ie the 3D \texttt{head-on}
in Table~\ref{tab:ID}) requires a particular refinement structure in
Cartesian coordinates. More specifically, we introduce five nested
rectangular regions with the innermost having the highest refinement
level, characterized by a volume of $90\times40\times40\,M_{\odot}^3
\simeq 133\times59\times59\,\mathrm{km^3}$) and containing both stars at
all times. Each side of outer refinement boxes having $(x,y,z)$ extents
in solar masses given by $(10, 10, 10)$, $(20, 30, 30)$, $(110, 30, 30)$,
and $(160, 180, 180)$.

Note that whenever importing initial data in \BHACns, we check the
L{\"o}hner refinement criterion to establish whether to refine or coarsen
the grid blocks and refill the initial data to the refined grids before
the evolution in order to reduce the error induced by prolongation and
restriction for the initial data. During the evolution, instead, we
evaluate the L{\"o}hner refinement criterion every $10$ iterations to
determine new refinement levels for all blocks.

Finally, for all simulations performed with \BHACns, the values of the
CFL factor and of the efficiency parameter depend on the coordinate
employed, so that $\mathcal{C}_{_{\rm CFL}}= 0.3, 0.3$, and $0.4$, while
$\xi_{\rm eff} = 1/10, 1$, and $1/50$ for for cylindrical, Cartesian, and
spherical coordinates, respectively. The only exceptions are represented
by the tests involving the 3D head-on collision of two neutron stars and
the 2D long-term evolution of the BNS remnant; in particular, for the
head-on test we employ a Cartesian coordinate and a CFL factor $0.3$,
while for the 2D long-term evolution the cylindrical coordinate and a CFL
factor of $0.25$ are used. In both cases, different values of $\xi_{\rm
  eff} = 1, 1/3, 1/10$ are employed to determine the optimal balance
between computational costs and accuracy (see discussion in
Secs.~\ref{sec:head-on} and \ref{sec:longtermBNS}).

\begin{table*}[ht!]
  \centering
  \footnotesize
  \begin{tabular}{lcccccccccc}
    \hline 
    \hline 
    Test & Dim. & Coords. & Spacetime & EOS & $N_{\mathrm{ref}}$ &
    $N_1 \!\times\! N_2 \!\times\! N_3$ & $(N_1 \!\times\! N_2 \!\times\! N_3)_{\mathrm{eff}}$ & $\Delta
    x^i_{\mathrm{min}}$ & $x^i_{\rm max}$ & ID \\
    & & & & &  & &  &  ($10^{-2}$) & & \\
    \hline  
    \texttt{BU0-cow}         & 1D &  spher. & fixed      & $\Gamma$-law & $1$  & $640$               & $640$                & [$7.81$]         & [$50$]        &\texttt{XNS}\\
    \texttt{BU0-dyn}         & 1D &  spher. & dynamical  & $\Gamma$-law & $1$  & $640$               & $640$                & [$9.38$]         & [$50$]        &\texttt{XNS}\\
    \texttt{migration}       & 2D &  spher. & dynamical  & $\Gamma$-law & $1$  & $640\!\times\!32$   & $640\!\times\!32$    & [$9.38, 5.33$]   & [$60,\pi/2$]  &\texttt{RNS}\\
    \texttt{magnetized-DRNS} & 2D &  spher. & dynamical  & $\Gamma$-law & $3$  & $64\!\times\!64$    & $256\!\times\!256$   & [$39.10, 0.61$]  & [$100,\pi/2$] &\texttt{XNS}\\
    \texttt{DD2RNS-mr}       & 2D &  cylin. & dynamical  & HSDD2        & $5$  & $32\!\times\!32$    & $512\!\times\!512$   & [$19.53, 19.53$] & [$+100,+100$] &\texttt{RNS}\\
    \texttt{DD2RNS-hr}       & 2D &  cylin. & dynamical  & HSDD2        & $5$  & $64\!\times\!64$    & $1024\!\times\!1024$ & [$9.77, 9.77$]   & [$+100,+100$] &\texttt{RNS}\\
    \texttt{head-on}         & 3D &  Cart.  & dynamical  & HSDD2        & $5$  & $128\!\times\!128\!\times\!64$  & $2048\!\times\!2048\!\times\!1024$ & [$19.53,19.53,19.53$]  & [$\pm200,\pm200,+200$] &\texttt{FUKA}\\
    \texttt{DD2BNS-HO@20ms}  & 2D &  cylin. & dynamical  & HSDD2        & $10$ & $16\!\times\!16$    & $8192\!\times\!8192$ & [$9.77, 9.77$]   & [$+800,+800$] &\FIL\\
    \texttt{DD2BNS-HO@50ms}  & 2D &  cylin. & dynamical  & HSDD2        & $10$ & $16\!\times\!16$    & $8192\!\times\!8192$ & [$9.77, 9.77$]   & [$+800,+800$] &\FIL\\
    \hline  
    \hline 
  \end{tabular}
  \caption{Summary of the simulations discussed in the paper and
    performed by \BHACns. The different columns report the name of the
    test, the number of spatial dimensions and the coordinates employed,
    the evolution of the spacetime, the EOS, the number of refinement
    levels ($N_{\rm ref}$), the number of cells on the coarsest level
    ($N_1 \times N_2 \times N_3$), the effective number of cells given
    the refinements ($(N_1 \times N_2 \times N_3)_{\rm eff}$), the
    finest-cell sizes in the first, second, and third dimension $(\Delta
    x^i_{\rm min})$; depending on the coordinate system the units are
    either solar masses or radians), the maximum coordinate in the first,
    second, and third dimension $x^{i}_{\rm max}$ (domain), and the code
    providing the initial data. Information on the simulations run by
    \FIL are reported in the main text.}
  \label{tab:ID}
\end{table*}

\begin{figure}
\includegraphics[width=0.49\textwidth]{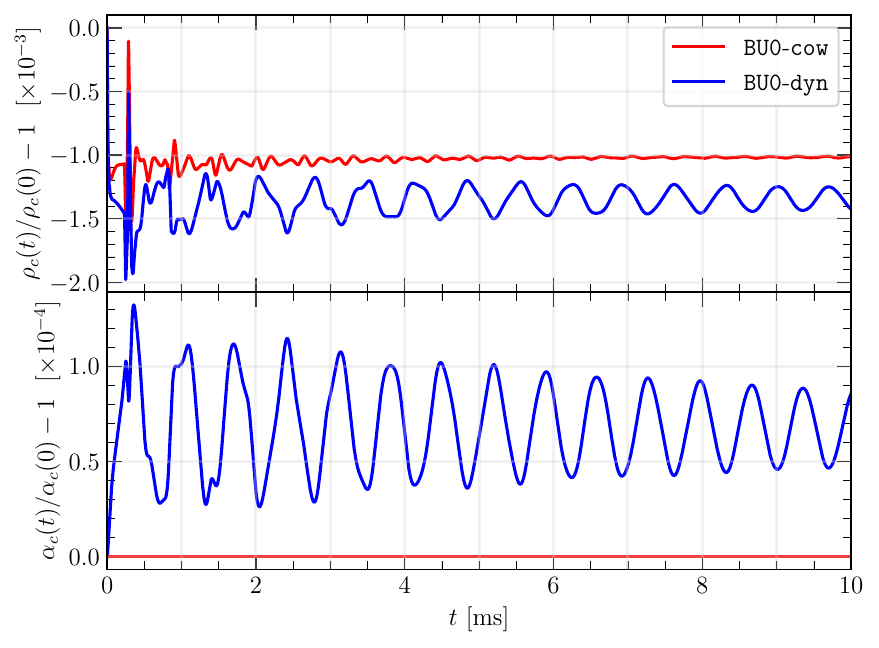}
\caption{\textit{Top panel:} Relative difference in the evolution of the
  central rest-mass density $\rho_c(t)$ when normalized to the initial
  one for two nonrotating stars when evolved in either the Cowling
  approximation (\texttt{BU0-cow}, red solid line) or with a dynamical
  spacetime within the CFC approximation (\texttt{BU0-dyn}, blue solid
  line). \textit{Bottom panel:} the same as in the top one but for the
  central value of the lapse function $\alpha_{c}(t)$.}
\label{fig:IG_1D_cfc_vs_cowling}
\end{figure}

\subsection{TOV star with an ideal-fluid EOS}
\label{sec:TOV_star}

We start our validation of \BHAC with a rather simple but complete test:
the long-term oscillation properties of a nonrotating star. This test was
first employed in a fully general-relativistic spacetime already in
Ref.~\cite{Font02c} and has been explored systematically in
Ref.~\cite{Dimmelmeier06}. Hence, we consider a nonrotating neutron-star
model with a polytropic EOS having $\Gamma = 2$ and $K = 100$, a
gravitational mass of $1.40\,{M_{\odot}}$, and a central rest-mass
density of $\rho_{c} = 1.28 \times 10^{-3}= 7.91 \times 10^{15}\,{\rm
  g~cm}^{-3}$. This model was first introduced in
Ref.~\cite{Stergioulas04} and is there referred to as ``BU0'', and is
also a reference model for the open-source code
\texttt{XNS}~\cite{Bucciantini2011, Pili2014}. Given the symmetry of the
problem and the flexible dimensionality of \BHACns, we carry out this
first test in 1D spherical coordinates and an ideal-fluid ($\Gamma$-law)
EOS, \ie $p = \rho \epsilon (\Gamma - 1)$, with $\Gamma = 2$. The
spacetime is either kept fixed in what is otherwise referred to as the
``Cowling approximation'', (test \texttt{BU0-cow}) or evolved with the
xCFC scheme (test \texttt{BU0-dyn}); in both cases the evolution is
carried out for $10\,\mathrm{ms}$.

The upper panel of Fig.~\ref{fig:IG_1D_cfc_vs_cowling} shows the relative
difference in the evolution of the central rest-mass density
$\rho_{c}(t)$ with respect to the initial rest-mass central density, \ie
$\rho_c(0)$, for both the models \texttt{BU0-cow} and \texttt{BU0-dyn}.
The lower panel of Fig.~\ref{fig:IG_1D_cfc_vs_cowling}, on the other
hand, shows the same but for the central value of the lapse function
$\alpha_{c}(t)$. Since no explicit perturbation is introduced in both
stars, the small oscillations are triggered by round-off errors and they
remain harmonic and small in amplitude (\ie $\lesssim 10^{-3}$) over the
time of the solution.

The evolutions in Fig.~\ref{fig:IG_1D_cfc_vs_cowling} report a well-known
behaviour (see, \eg \cite{Font02c, Dimmelmeier06, Radice2013c}), namely,
that the oscillations are more rapidly damped in the case of a fixed
spacetime, simply because the dynamical coupling between the evolution of
matter and the gravitational field is broken and a larger amount of mass
is lost at the surface. Stated differently, in the Cowling approximation
the gravitational fields cannot react to the local under- or
overdensities caused by oscillations and matter is more easily lost from
the stellar surface at each oscillation. When the spacetime is
evolved, on the other hand, the amplitude of the oscillations is damped
because of a small but nonzero numerical bulk
viscosity~\cite{CerdaDuran2010, Chabanov2023}.

\begin{figure}
\includegraphics[width=0.45\textwidth]{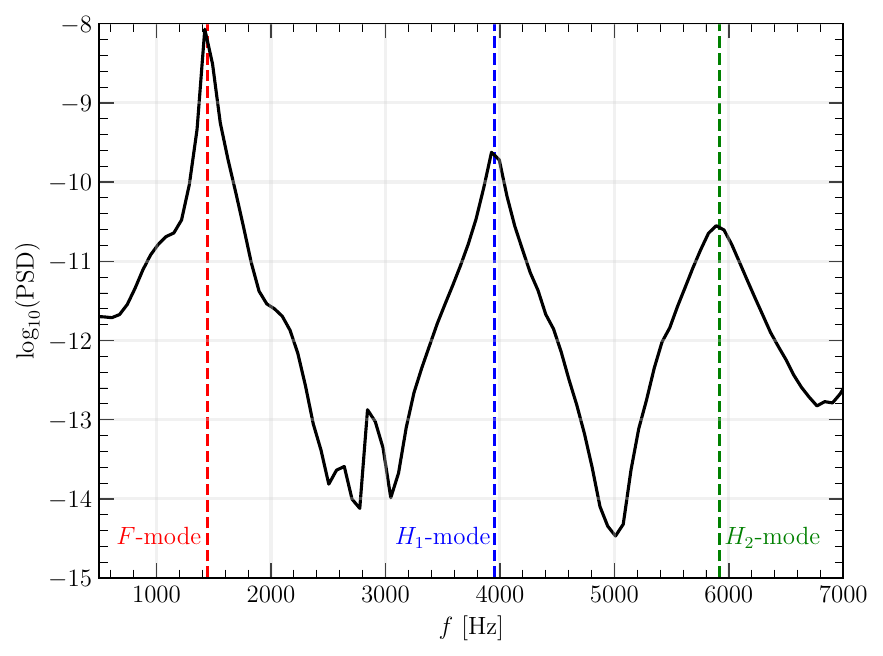}
\caption{Power spectral density (PSD) of the evolution of the normalized
  central rest-mass density $\rho_c(t)/\rho_c(0)$ of the \texttt{BU0-dyn}
  test computed over a timescale of $10\,\mathrm{ms}$ as in the top panel
  of Fig.~\ref{fig:IG_1D_cfc_vs_cowling}. Two clear peaks are visible in
  the PSD and show a very good match with the expected eigenfrequencies
  of the fundamental mode ($F$-mode, red dashed line), of its first
  overtone ($H_1$-mode; blue dashed line) and second overtone
    ($H_2$-mode; green dashed line) computed from perturbative
    studies~\cite{Font02c}}.
\label{fig:IG_1D_cfc_fft}
\end{figure}

What matters most in this test is that the frequencies of the numerical
oscillations match the expected oscillation eigenfrequencies computed
perturbatively, either in a fixed or in a dynamical spacetime. To this
scope, Fig.~\ref{fig:IG_1D_cfc_fft} reports the power spectral density
(PSD) of the Fourier transform of the function $\rho_{c}(t)$ over the
$10\,\mathrm{ms}$ evolution of model \texttt{BU0-dyn} and compares it
with the perturbative frequencies of the fundamental radial-oscillation
mode ($F$-mode), its first overtone ($H_1$) and second overtone
  ($H_2$)~\cite{Font02c}.  The relative difference between the two
  frequencies is $-0.07\%$ for the $F$-mode, $-0.43\%$ for the
  $H_1$-mode, and $-0.61\%$ for the $H_2$-mode, respectively, with the
numerical mode being systematically smaller, as expected from a nonlinear
solution in a linear regime. Overall, the high accuracy of these results
provide us with the first evidence of the correct implementation of the
CFC approximation in \BHACns.

\subsection{Migration test}
\label{sec:test_mt}

Stepping up in complexity, we now consider a test that simulates a fully
nonlinear scenario in which both the field and the matter variables
undergo very rapid changes. The test in question, which is commonly
referred to as the ``\texttt{migration} test'' was first introduced in
Ref.~\cite{Font02c} and has since been employed to test a variety of
codes~\cite{Cordero2009, Bucciantini2011, Radice2013c}. In essence, this
test studies the evolution of a nonrotating neutron star placed on the
unstable branch of equilibrium configuration and which is triggered to
``migrate'' on the stable branch where it will find a stable
configuration with the same rest mass after undergoing a series of
large-amplitude oscillations. In this process, the star essentially
expands very rapidly, converting its binding energy into kinetic energy,
and then, via shock-heating, into internal energy.

Since this is purely a numerical test, we choose the neutron star to have
a central rest-mass density of $\rho_c = 7.993 \times 10^{-3}\simeq
4.937\times 10^{15}\,{\rm g~cm}^{-3}$, and employ a polytropic EOS with
$\Gamma = 2$ and $K = 100$, thus leading to an initial radius $R =
4.06\,M_{\odot}=6.29\,\mathrm{km}$. The evolution, on the other hand, is
carried out with an ideal-fluid EOS with the same adiabatic index. The
stellar model is then evolved in 2D employing spherical polar coordinates
within a dynamical spacetime and its dynamics compared with that obtained
with \FILns.

\begin{figure}
\center
\includegraphics[width=0.45\textwidth]{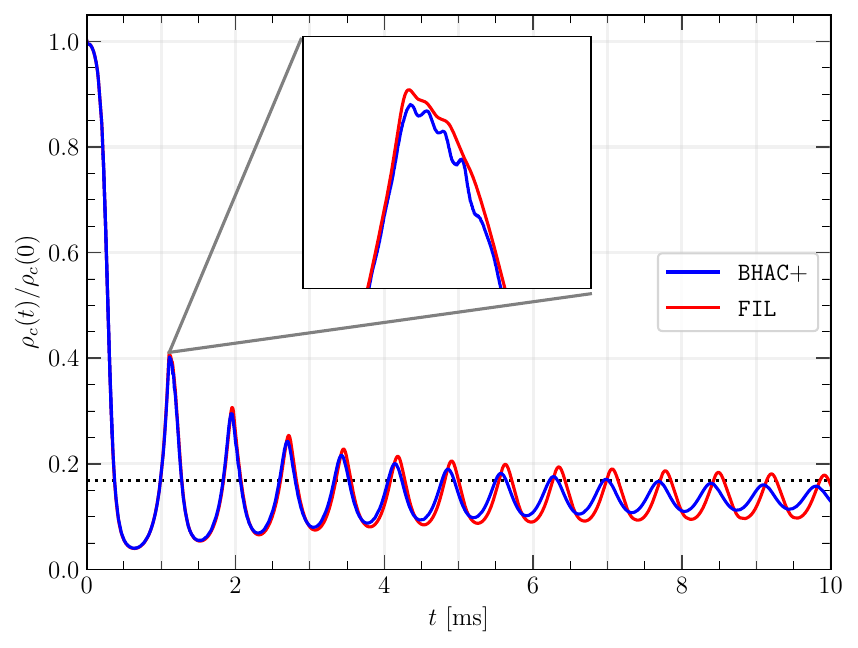}
\caption{Evolution of the central rest-mass density normalized to the
  initial value in the \texttt{migration} test. Reported with solid lines
  of different color are the evolution by \BHAC (blue line) and by \FIL
  (red line). Note the excellent agreement especially over the first few
  oscillations; the black dotted line represents the central rest-mass
  density of the star on the stable branch having the same gravitational
  mass, which is higher than the asymptotic solutions since it does not
  account for the matter lost in the nonlinear shocks at the stellar
  surface.}
\label{fig:IG_SU2D}
\end{figure}

\begin{figure*}
\centering
\includegraphics[width=0.52\textwidth]{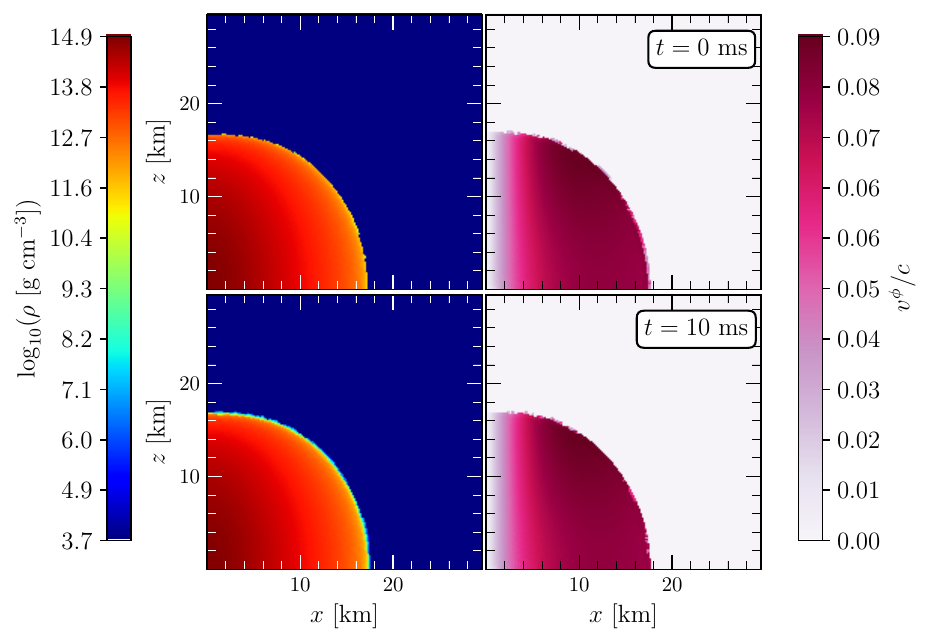}
\hspace{0.2cm}
\includegraphics[width=0.46\textwidth]{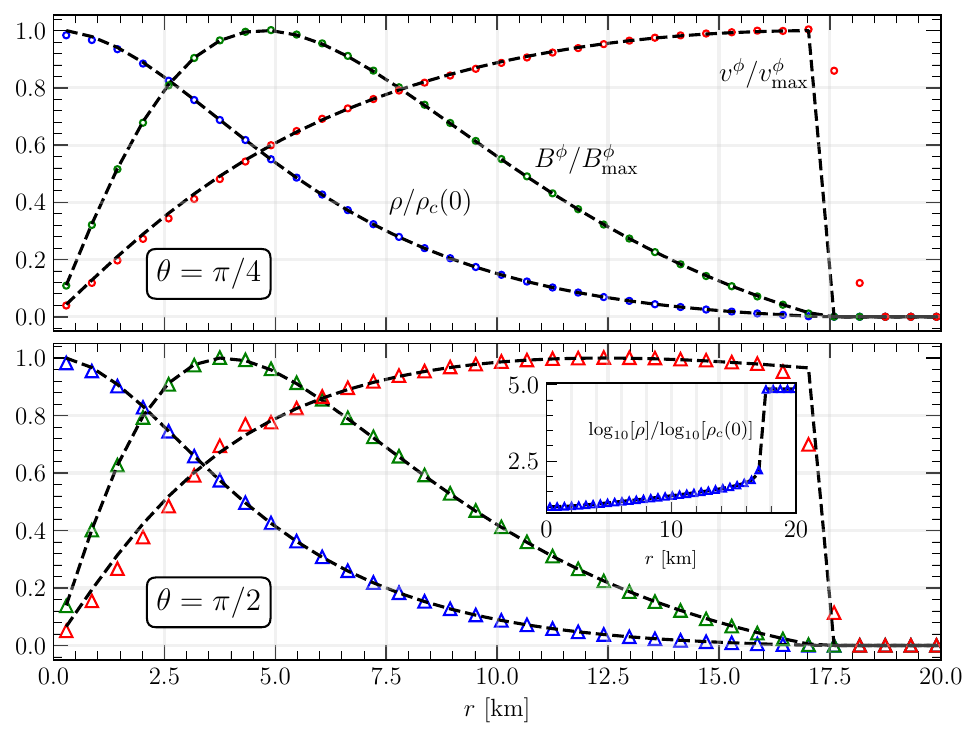}
\caption{\textit{Left panel:} 2D distributions of the rest-mass density
  (left column) and of the rotational velocity (right column) of a
  magnetized and differentially rotating star (test
  \texttt{magnetized-DRNS} in Table~\ref{tab:ID}). The top row refers to
  the initial time $t=0$, while the bottom row to the final time
  $t=10\,\mathrm{ms}$, corresponding to about eight rotational periods,
  and showing a very good preservation of the axisymmetric
  equilibrium. \textit{Right panel:} indicated with different symbols are
  the 1D profiles of the rest-mass density, of the angular velocity, and
  of the toroidal magnetic field, all normalized to their maximum values
  for the same test in the left panel. The top row displays the profiles
  at a polar angle $\theta = \pi/4$, while the bottom row shows them on
  the equatorial plane $\theta = \pi/2$. For all quantities, the dashed
  lines represent the initial profiles, which are well preserved even
  after eight rotational periods. The inset in the bottom row
    reports $\mathrm{log}_{10}[\rho]/\mathrm{log}_{10}[\rho_c(0)]$ and
    shows that the surface of the star is captured with a couple of cells
    only.}
\label{fig:MDRNS_2dslices}
\end{figure*}

Figure~\ref{fig:IG_SU2D} illustrates the evolution of the central
rest-mass density normalized by its initial value at $t = 0$, while the
black dotted line represents the central rest-mass density of the neutron
star on the stable branch with $\rho_c = 1.346 \times 10^{-3}$ having the
same gravitational mass as the initial model (this value is higher than
the asymptotic solutions since it does not account for the matter lost in
the nonlinear shocks at the stellar surface). Overall, our results are
qualitatively consistent with previous studies, either in full general
relativity~\cite{Font02c, Radice2013c}, or employing the CFC
approximation~\cite{Cordero2009, Bucciantini2011}, and exhibit the
well-know behaviour in terms of peak amplitudes, density at the first and
second maxima, the non-harmonic nature of the density oscillations,
etc. However, for a more quantitative comparison, we present in
Fig.~\ref{fig:IG_SU2D} also a direct comparison of the corresponding
evolution carried out by \FIL in full general relativity and with very
similar spatial resolution. Notwithstanding the intrinsic approximations
associated with the CFC approach, the similarities between the two
curves, especially in the most nonlinear part of the evolution (\ie $t
\lesssim 2\,{\rm ms}$) is quite remarkable; the similarities between the
two evolutions persist up to $t \lesssim 5\,{\rm ms}$, after which the
more dissipative features of the CFC approximation appear and phase
differences emerge in the evolution. We should recall, in fact, that, in
addition to the less accurate spacetime evolution, \BHAC utilizes a
second-order accurate finite-volume scheme for the solution of the GRMHD
equations, while \FIL employs a fourth-order accurate -- and hence less
diffusive -- finite-difference method. Overall, however, also this
\texttt{migration} test provides an important validation of the correct
implementation of the CFC solver in a 2D scenario.

\subsection{Magnetized and differentially rotating star}

All of the tests presented so far referred to configurations with a zero
magnetic field. In order to validate the ability of \BHAC to properly
solve the GRMHD equations in a dynamical spacetime, we consider the
evolution of a magnetized and differentially rotating
star~\cite{Bucciantini2011, Cheong2021}. To this scope, we again use
\texttt{XNS}~\cite{Bucciantini2011} to generate a self-consistent
magnetized star with a purely toroidal magnetic field and in differential
rotation. In particular, the initial stellar model was modeled as
following the $j$-constant rotation law~\cite{Komatsu89b} with central
angular velocity $\Omega_c = 2.575 \times 10^{-2}$ and
differential-rotation parameter $A^2 = 70$, and a polytropic EOS with
$\Gamma = 2$ and $K = 100$ (this test is referred to as
\texttt{magnetized-DRNS} in Table~\ref{tab:ID}). The resulting initial
central rest-mass density is $\rho_c(0) = 1.28 \times 10^{-3} = 7.91
\times 10^{15}\,{\rm g~cm}^{-3}$ and we prescribe the magnetic-field
strength $B := \sqrt{B_i B^i}$ with the law~\cite{Kiuchi2008}
\begin{equation}
  B :=\left\{
  \begin{array}{lr}
    {K_m (\alpha^2 \varpi^2 \rho h)^{m}}/({\alpha \varpi})\,, &\text{for}~\rho > 10^{-9} \\
    0\,, &\text{for}~\rho \le 10^{-9}
\end{array}\right.
\end{equation}
where $\varpi := \psi^2 r \sin\theta$ is the generalized cylindrical
radius, with $r$ and $\theta$ being the spherical radial and polar
coordinates; in practice, we set $m=1$ and $K_m = 3$. As remarked in
Refs.~\cite{Bucciantini2011, Cheong2021}, the magnetic field in this star
reaches a maximum value of $\sim 5 \times 10^{17}\,\mathrm{G}$, thus
accounting for $\sim 10\%$ of the total internal energy of the star and
providing a non-negligible change in the underlying equilibrium.

\begin{figure*}
\centering
\includegraphics[width=0.45\textwidth]{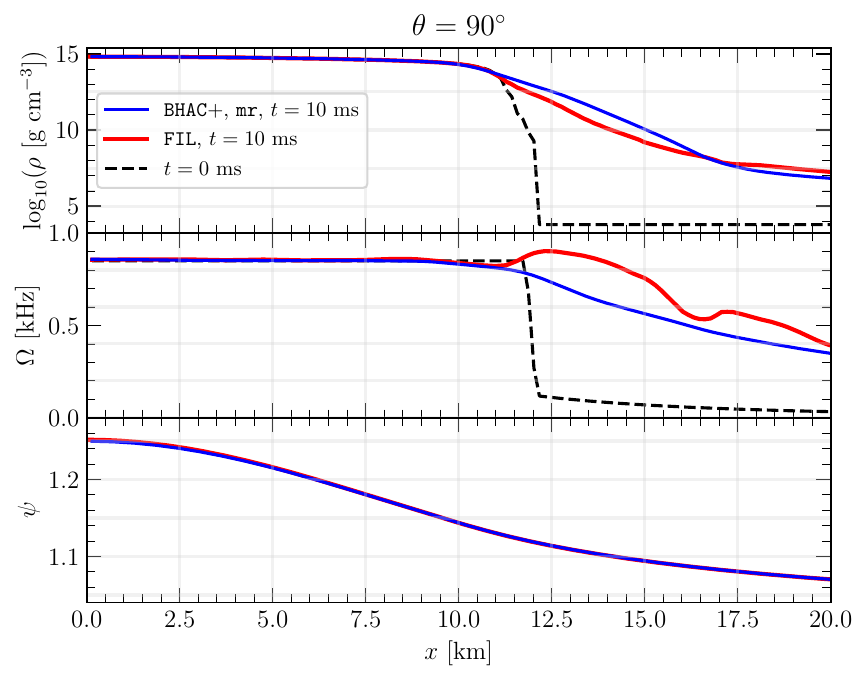}
\hspace{1.0cm}
\includegraphics[width=0.45\textwidth]{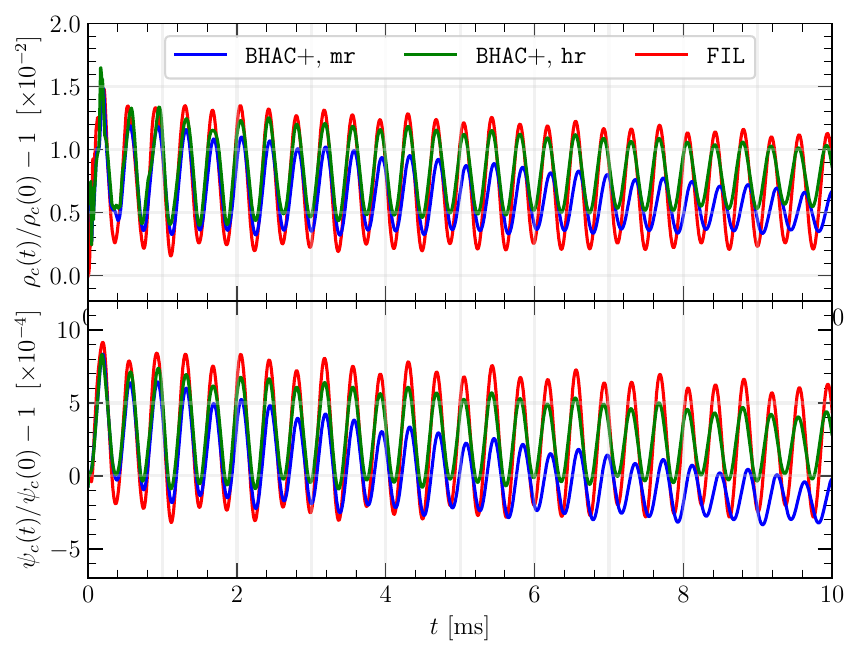}
\caption{\textit{Left panel:} 1D profiles at $\theta = 90^{\circ}$ and at
  $t=10\,\mathrm{ms}$ of the rest-mass density (top row), of the angular
  velocity (middle row), and of the conformal factor (bottom row) for a
  rapidly and uniformly rotating neutron star (test \texttt{DD2RNS-mr} in
  Table~\ref{tab:ID}). Solid lines of different color refer to the
  evolutions carried out by \BHAC (blue line) or \FIL (red line) showing
  a very good preservation of the equilibrium in the high-density regions
  of the star. The deviations from the initial profiles at
  $t=0\,\mathrm{ms}$ (black dashed lines) are comparable in the two codes
  and are typical of simulations of rapidly rotating stars (note the
  logarithmic scale employed here).  \textit{Right panel:} Evolution of
  the relative difference in the central rest-mass density (top row) and
  of the conformal factor (bottom row) for the same model reported in the
  left panel and the same convention for the line types. In addition,
  each row contains also the data of a high-resolution simulation
  (\texttt{DD2RNS-hr}, green solid line), highlighting how the
  differences with \FIL can be decreased by a higher spatial resolution.}
\label{fig:DD2RNS_1dslices}
\end{figure*}

Once the initial stellar model is imported in \BHACns, we evolved the
system in 2D using spherical coordinates and the same ideal-fluid EOS
employed in the previous test for a duration of $10 \, \mathrm{ms}$. The
left panel of Fig.~\ref{fig:MDRNS_2dslices} shows the 2D slices of the
rest-mass density and of the rotational velocity, $v^{\phi}$, at two
different times, $t=0$ (upper part) and $t=10\,\mathrm{ms}$ (lower
part). Clearly, a direct and qualitative comparison of the two 2D slices
shows the ability of the code to retain an accurate description of the
stellar model over more than eight spinning periods. The right panel of
Fig.~\ref{fig:MDRNS_2dslices} shows instead a more quantitative
comparison of the radial profiles of the rest-mass density, the linear
rotational velocity $v^{\phi}$, and toroidal magnetic field $B^{\phi}$ at
$t=0$ and $t=10\,\mathrm{ms}$ which are normalized by their maximum
values. Furthermore, the upper part of the panel refers to the diagonal
direction ($\theta = \pi/4$), while the lower panel to the equatorial one
($\theta = \pi/2$). We note that there on both angles there are minor
distortions in the rotational velocity $v^{\phi}$ around $r \approx 17$
$\mathrm{km}$, where the low-density atmosphere interfaces with the
high-density neutron star. The sharp gradient introduced by the stellar
surface oscillates as a result of the round-off perturbations exhibiting
a behavior consistent with the findings of
Ref.~\cite{Bucciantini2011}. Overall, the results of this test further
demonstrate that \BHAC is capable of stably simulating a rapidly
configuration stellar configuration with a strong magnetic field and over
several rotation periods.

\subsection{Rapidly and uniformly rotating star with tabulated EOS}

Next, we validate our new code by evolving a rapidly and uniformly
rotating neutron star with a rotation rate close to the mass-shedding
limit and described by a tabulated, finite-temperature EOS, specifically
the HSDD2 EOS~\cite{Hempel2010}. Our initial data is computed as an
axisymmetric equilibrium model using the \texttt{RNS}
code~\cite{Stergioulas95} with angular velocity $\Omega = 2.633 \times
10^{-2} = 850.85 \, 2\pi\,{\rm Hz}$ and is assumed to be in a neutrinoless
$\beta$-equilibrium state with $T=T_{\mathrm{min}}$. The evolution in
\BHAC is performed in 2D with cylindrical coordinates and $z$-symmetry
while, at the same time, we carry out an analogous evolution with \FIL in
Cartesian coordinates with the same resolution over the star (this is the
test \texttt{DD2RNS-mr} in Table~\ref{tab:ID}). To quantify the
resolution dependence of \BHACns, we perform an additional simulation
with \BHAC having a resolution that is twice that used in \FIL (this is
the test \texttt{DD2RNS-hr} in Table~\ref{tab:ID}).

The left panel of Fig.~\ref{fig:DD2RNS_1dslices} illustrates the profiles
on the equatorial plane (\ie $z=0$) of the rest-mass density $\rho$ (top
panel), of the angular velocity $\Omega$ (middle panel), and of conformal
factor $\psi$ (bottom panel) at the initial time (black dotted line) and
at $t=10\,\mathrm{ms}$, both for \BHAC (blue solid line, case
\texttt{DD2RNS-mr}) and \FIL (red solid line). Remarkably, after eight
rotation periods, all the matter quantities in the stellar interiors (\ie
$x \lesssim 12\,\mathrm{km}$) are well preserved, with only small
deviations from the initial data despite the very extreme properties of
the stellar model. This is true both for the data obtained with \BHAC and
with \FILns; an even better agreement is found in the conformal factor,
where the relative differences are less than $\simeq 0.15\%$.

The right panel of Fig.~\ref{fig:DD2RNS_1dslices}, on the other hand,
reports the relative differences in the evolution of the central
rest-mass density $\rho_c$ and of the conformal factor $\psi_c$ when
compared to their initial values. Note also that in the case of the
simulations carried out by \BHACns, we report evolutions with two
different resolutions. This shows that as the resolution of \BHAC is
increased, the differences to the evolution in \FIL decreases and the
small de-phasing observed in the case of the medium-resolution simulation
decreases significantly. The oscillations in $\rho_c$ and $\psi_c$ show
relative variations in the high-resolution simulations that are less than
$10^{-2}$ and $10^{-4}$, respectively. Note that we have performed
  an extra simulation with medium resolution and with $\xi_{\mathrm{eff}}
  = 1$ that is not shown in the right panel of
  Fig.~\ref{fig:DD2RNS_1dslices}. As expected, in this case we find a
  damping timescale that is longer than that measured in the case of
  \texttt{DD2RNS-hr} with $\xi_{\mathrm{eff}} = 1/10$, and a solution
  that is less diffusive than in both of the cases of \texttt{DD2RNS-hr} and
  \texttt{DD2RNS-mr}.

Overall, bearing in mind that \FIL uses high-order methods and the full
evolution of the spacetime, the agreement with \BHACns, already at
comparatively small resolutions and in simulating a rather challenging
stellar model, confirms the ability of \BHAC and of the CFC approximation
to accurately reproduce in 2D results from a full 3D numerical-relativity
code. In the following section we will demonstrate that this is also the 
case in full 3D simulations.

\subsection{Head-on collision of two neutron stars}
\label{sec:head-on}

We next discuss the head-on collision of two neutron stars as a 3D test
to validate the full implementation of the set of equations and explore
conditions of spacetime curvature and matter dynamics that are very
similar to those encountered in a binary merger from quasi-circular
orbits~\cite{Musolino2023}, but that can be tested at a fraction of the
computational cost (this test is indicated as \texttt{head-on} in
Table~\ref{tab:ID}). Indeed, the head-on collision of two stars has a
long history and has been in the past employed to actually study the
dynamics of critical phenomena~\cite{Kellermann:10} or the formation of
black holes for ultrarelativistic initial speeds~\cite{East2012,
  Rezzolla2013}. Furthermore, because of the minimal influence of
gravitational waves, this scenario is also particularly suited for
assessing codes utilizing the CFC approximation and allows us to compare
once again the solutions obtained with \FIL and \BHACns.

The initial data of \FIL is generated using the \texttt{FUKA}
code~\cite{Grandclement09, Papenfort2021b, Tootle2023a}, which computes
the initial data timeslice by solving the eXtended Conformal Thin
Sandwhich (XCTS) system of equations~\cite{Pfeiffer:2005,
  Papenfort2021b}. The initial data is obtained by first computing the
isolated 3D solutions of the stars prior to constructing a spacetime
representing the binary system. However, unlike the implementation
discussed in Refs.~\cite{Papenfort2021b, Tootle2023a}, we approximate the
solution by superimposing two isolated solutions and re-solving the XCTS
constraint equations where, however, some care must be taken as we will
discuss shortly.

The initial guess of the head-on is generated by superimposing the
isolated stellar solutions such that, for a given spacetime or source
field $X$, the initial guess in the binary is constructed
as~\cite{Tootle2023a, Tootle2024a}
\begin{align}
  X_{\rm bin} \left( \boldsymbol{x} \right) &:= 
  \Xi + \kappa_1 \left(X_1\left( \boldsymbol{\hat{x}_1} \right)\!-\!\Xi\right)
      + \kappa_2 \left(X_2\left( \boldsymbol{\hat{x}_2} \right)\!-\!\Xi\right) \,, \\
  \kappa_{1,2} &:= 
    \exp \left[ - \left( \frac{r_{1,2}}{d_0/2} \right)^4 \right] \,, \\
  \boldsymbol{\hat{x}}_{1,2} &:= \boldsymbol{x} - \boldsymbol{{x}}_{c1,c2} \,, 
\end{align}
where $\Xi$ is the asymptotic value for a given field (\eg $\psi = \alpha
= 1$, $\beta^i = 0$, etc), $d_0$ is the initial separation,
$\boldsymbol{{x}}_{c1,c2}$ are the location of the neutron-star centers,
and $\kappa_{1,2}$ represent the ``decay parameters'' centered about the
respective neutron-star solution, such that the solution is exactly the
isolated solution near the neutron star and then decays to flat spacetime
further away. The decay behaviour of the solutions is controlled by the
$4^{\rm th}$ power in the exponential, while the decay distance is
controlled by the weight factor $d_0$. This approach is analogous to that
employed in Ref.~\cite{Helfer:2021brt} for the head-on collision of boson
stars and was inspired by previous works~\cite{Lovelace2008c,
  Foucart2008}, though the application was focused on fixing background
metric quantities instead of generating an initial guess for obtaining an
initial-data solution.

\begin{figure}
\center
\includegraphics[width=0.45\textwidth]{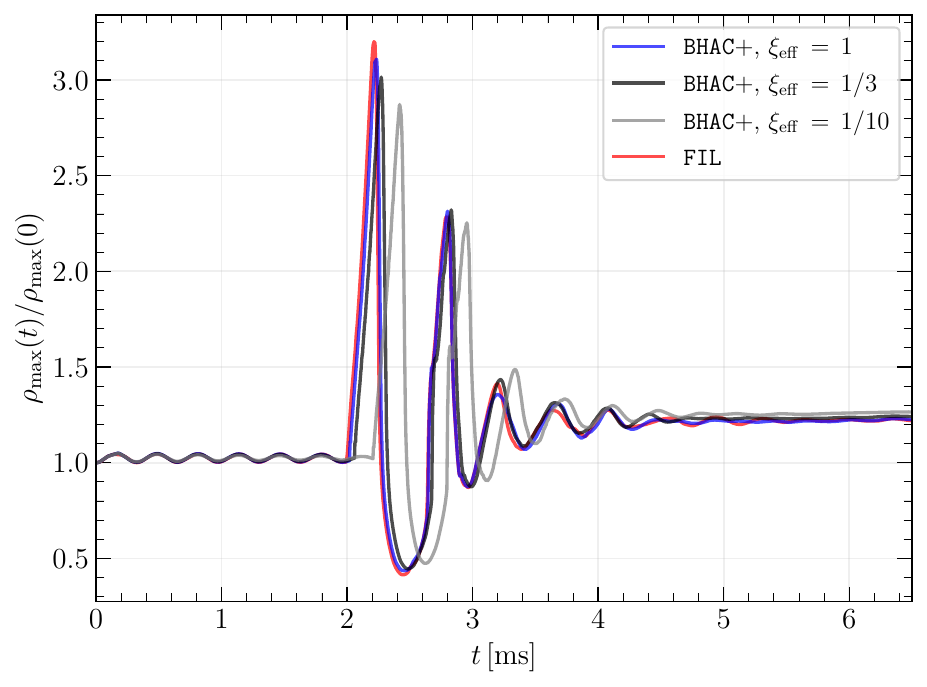}
\caption{Evolution of the maximum rest-mass density normalized to the
  initial value in the 3D \texttt{head-on} collision test. The red line
  represents the result of \FIL, while the blue line shows the evolution
  in \BHACns. Also shown with a black and gray solid line are the
  evolutions with different efficiency ratios, namely $\xi_{\rm eff} =
  1/3$ and $1/10$, respectively. The blue solid line is obtained with
  $\xi_{\rm eff} = 1$ and is obviously the closest to the \FIL
  evolution.}
\label{fig:headon_rhomax_compare}
\end{figure}

Since the initial timeslice is not (quasi-)stationary and we no longer
have a notion of conservation along fluid-lines, we are not able to
strictly enforce hydrostatic equilibrium by solving the Euler equation.
Instead, we adopt an approach similar to that used in
Ref.~\cite{Papenfort2021b}, where we relax this constraint and simply
rescale the fluid quantities by a constant fixed by enforcing a fixed
rest-mass. Thus, the fluid description of each neutron star will
systematically scale as a function of the Lorentz factor $W$ due to the
presence of the companion object. For this reason, we set the initial
separation between the two stars to $d_0 = 60\,M_{\odot}\simeq 89.4\,{\rm
  km}$, such that the solutions are minimally rescaled while still
resulting in a computationally efficient setup. It is important to note
that relaxing the hydrostatic equilibrium is also a necessary step to
obtain eccentricity reduced initial data, the effects of which have been
discussed previously~\cite{Tichy2019, Papenfort2021b}.

To further test the interfacing of \FIL with \BHACns, the initial data
from \texttt{FUKA} is first imported from \FIL and then ``handed-off'' to
\BHAC so that the two codes have initial data that is equivalent to the
one they exchange in a typical HO situation. The initial velocity of two
neutron stars is set to zero, and their mass is set so as to avoid
black-hole formation, \ie they have an ADM mass $M_{_{\rm ADM}} =
0.91\,M_{\odot}$~\cite{Musolino2023}, and are described by the HSDD2
EOS~\cite{Hempel2010}. The finest refinement level of the two codes
contain both of the neutron stars, have a grid resolution of $\simeq
0.2\,M_{\odot}\simeq 298\,\mathrm{m}$; for simplicity and a closer
comparison, both the resolution and the grid structure is not varied
during the evolution.

Figure~\ref{fig:headon_rhomax_compare} reports the evolution of the
maximum rest-mass density normalized to its initial value as obtained by
\BHAC (blue solid line) and by \FIL (red solid line). Note the very good
agreement despite the very different set of field equations solved. In
particular, it is remarkable that not only the time of the collision (\ie
when the central rest-mass density deviates most significantly from its
initial value\footnote{Note that a less frequent spacetime update has the
consequence that the spacetime evolves ``less rapidly'' and this
obviously leads to a systematic delay in the time of collision.}) differ
by $\lesssim 1.3\%$, but also that the maximum and minimum changes in the
maximum rest-mass density are very similar and differ by $\lesssim 4.9\%$
at most, while the differences in the asymptotic equilibrium values of
the collision remain below $\simeq 0.4\%$. Note also that the variations
in the fluid variables in this case are much more severe and extreme than
what simulated in the case of the migration test (compare
Fig.~\ref{fig:headon_rhomax_compare} with Fig.~\ref{fig:IG_SU2D}).

Also shown in Fig.~\ref{fig:headon_rhomax_compare} are examples of
evolutions with different efficiency ratios and hence different
computational efficiency. More specifically, while the blue solid line
refers to $\xi_{\rm eff} = 1$, the black and gray solid lines refer to
$\xi_{\rm eff} = 1/3$ and $1/10$, respectively. Note that, as expected,
the comparison with the \FIL evolution are worse in these cases, but also
that the differences remain $\lesssim 10\%$ in the maximum variation of
the central density, while the actual fundamental frequency of
oscillation or the final central rest-mass density of the collision remnant
differ by $\lesssim 3\%$. These differences -- which are measured in the
most extreme conditions of spacetime curvature expected in BNS mergers
and are therefore to be taken really as upper limits -- need to be
contrasted with the corresponding gain in computational costs. More
specifically, given similar resolution between two codes, considering
that \BHAC is about $3.5~(4.6)~[6.3]$ times faster than \FIL for
comparable resolutions when setting $\xi_{\rm eff} = 1~(1/3)~[1/10]$, it
becomes clear that a systematic error of a few percent can be tolerated
over timescales of seconds when it comes with a gain of about a factor
four to six in computational costs. Furthermore, additional gains can
come from a better coordinate system and mesh refinement structure, by
the use of even smaller values of $\xi_{\rm eff}$ at later times as the
spacetime dynamics is much less severe, and, more importantly, by the
considerable difference in the CFL constraint when considering the sound
speed in place of the speed of light. Note that it is not difficult to
show analytically that the computational gain $\bar{\gamma}$, \ie the
ratio of operations in a fully general-relativistic code (\eg \FIL) and
of constraint-solving code (\eg \BHAC), is $\bar{\gamma}=(1+\xi_{\rm
  eff}c_s/c)/[(1+\xi_{\rm eff})c_s/c]$. Hence, $\bar{\gamma} \to c/c_s$
in the limit of $\xi_{\rm eff} \to 0$. Furthermore, because $\xi_{\rm
  eff}$ can be decreased with increasing spatial resolution, the
computational gain actually increases when performing simulations with
higher resolutions.

\begin{figure*}
\center
\includegraphics[width=0.49\textwidth]{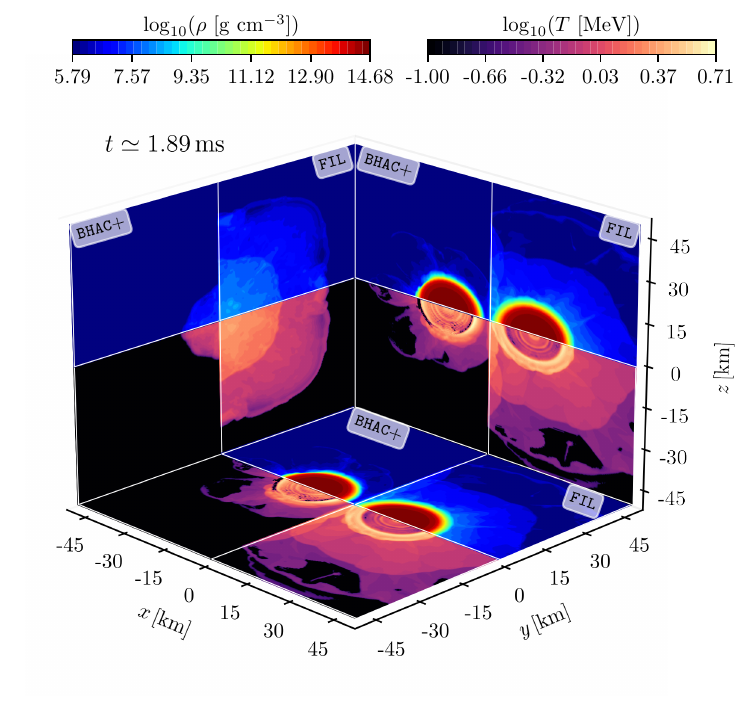}
\hskip  0.25cm
\includegraphics[width=0.49\textwidth]{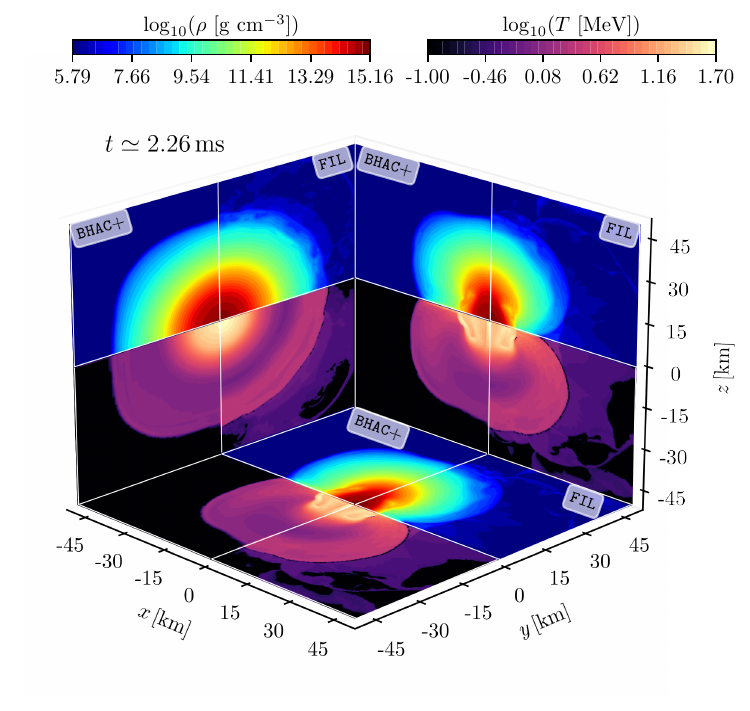}
\includegraphics[width=0.49\textwidth]{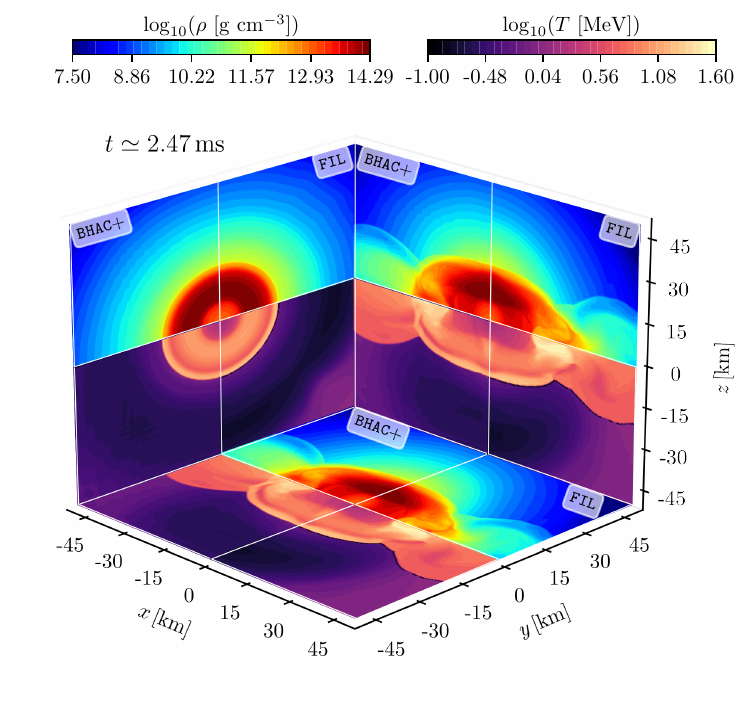}
\hskip  0.25cm
\includegraphics[width=0.49\textwidth]{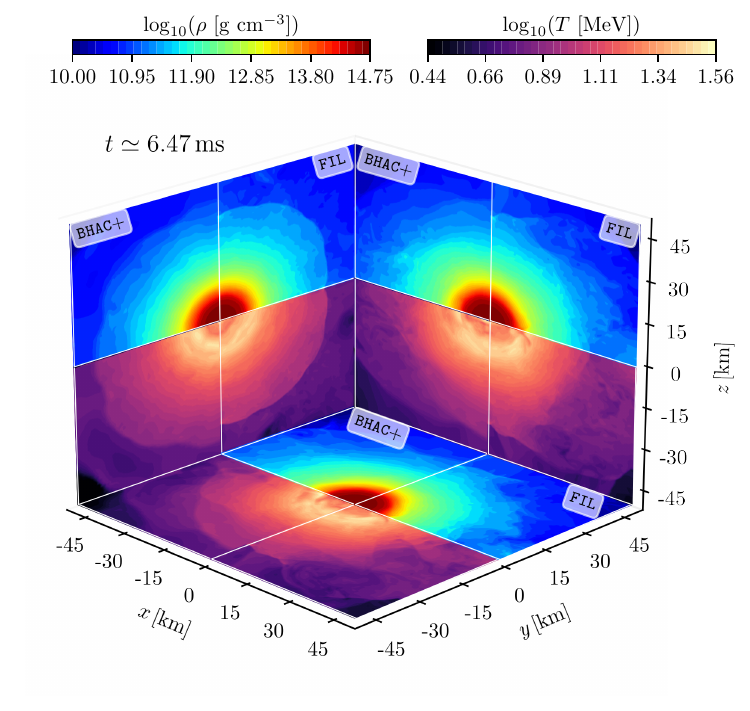}
\caption{2D comparison on the principal planes of the rest-mass density
  and temperature between in the head-on collision of two neutron stars
  (test \texttt{head-on}) as computed by \BHAC (left part of each plane)
  and \FIL (right part of each plane). In each column, the top part of
  each panel reports the rest-mass density, while the bottom part shows
  the temperature. Finally, the four panels refer to four different
  times, namely, the instant when the two neutron stars start colliding
  (top-left panel, $t\simeq 1.89 \,{\rm ms}$), that of the maximum
  compression (top-right panel, $t\simeq 2.26 \,{\rm ms}$), that of the
  minimum compression (bottom-left panel, $t\simeq 2.47 \,{\rm ms}$), and
  that of the late-time evolution (bottom-right panel, $t\simeq 6.47
  \,{\rm ms}$). Note that the colorbars are different in the four panels
  and the use of negative-$z$ regions is done for visualization purposes
  only since the simulations actually employ a symmetry across the $z=0$
  plane. Finally, note the very good agreement between the two evolutions
  despite the difference in coordinate systems, truncation order in the
  solution of the GRMHD equations, and different treatment of the
  spacetime evolution.}
\label{fig:headon_comparison_3d}
\end{figure*}

Figure~\ref{fig:headon_comparison_3d} offers a comprehensive comparison
of the 2D rest-mass density and temperature distributions of \BHAC
(depicted in the left part of each principal plane) and \FIL (depicted in
the right part of each principal plane). The different panels refer to
different and representative times during the collision and the top-left
panel in Fig.~\ref{fig:headon_comparison_3d}, in particular, reports the
instant when the two neutron stars start colliding at $t\simeq 1.89
\,{\rm ms}$. The comparison reveals a high degree of similarity between
the two codes, although small differences do emerge. In particular, and
as expected, the atmosphere surrounding \FIL is hotter and denser than
that of \BHACns. This discrepancy is mostly attributed to the different
order at which the GRMHD equations are solved in the two codes with
high-order schemes being normally more sensitive to small shocks at the
stellar surface and hence to very small mass losses [see, \eg
  Refs.~\cite{Radice2012a, Radice2013c, Guercilena17} for a
  discussion]. Note also that in this pre-merger phase both codes suffer
from small failures in the temperature near the stellar surface and once
again these are produced by the small rest-mass density fluctuations
near the surface, which, in turn, are amplified by the high-power
dependence of the temperature on these oscillations; when comparing the
behaviour of the internal specific energy, in fact, these oscillations
are essentially absent. The top-right panel of
Fig.~\ref{fig:headon_comparison_3d} shows instead the same quantities at
the instant of maximum compression at $t\simeq 2.26 \,{\rm ms}$ (see also
the first peak in Fig.~\ref{fig:headon_rhomax_compare}). At this time,
matter experiences extreme compression in the $x$-axis direction (the
collision is along the $x$-axis), leading to a peak temperature of
approximately $50 \, {\rm MeV}$ in the central region as a result of the
collision of the two strong shocks fronts. The rest-mass density and
temperature profiles in both codes exhibit striking similarity and some
differences appear only in the very low-density regions of the \FIL
evolution, which is absent in the \BHAC results. The bottom-left panel
refers instead to the instant of minimum compression at $t\simeq 2.47
\,{\rm ms}$ (see also the first minimum in
Fig.~\ref{fig:headon_rhomax_compare}), and shows that as a result of the
bounce and change of sign in the bulk linear momentum, matter is expelled
back in the $x$-axis direction producing a strong reverse shock at the
surface of the merged object, so that the remnant has a low-density,
low-temperature core produced by the induced rarefaction wave. Finally,
the bottom-right panel shows to the stable remnant at $t\simeq 6.47
\,{\rm ms}$, which exhibits a high-temperature mantle relative to the
shocked material and a comparatively cooler core.

In summary, the remarkably good agreement between the two fully 3D
evolutions despite the difference in coordinate systems, truncation order
in the solution of the GRMHD equations, and different treatment of the
spacetime evolution\footnote{Similar level of differences can be found
also when comparing the evolution of the same full numerical-relativity
code with slightly different hydrodynamical treatments (see,
\eg~\cite{Guercilena17}).} provides very convincing evidence for the
ability of the CFC approximation to effectively model gravitational
effects, particularly when the non-diagonal terms of the spatial metric
tensor in the system are anticipated to be negligible, as it is the case
during the free-fall stage preceding the collision.

\begin{figure*}
\center
\includegraphics[width=0.47\textwidth]{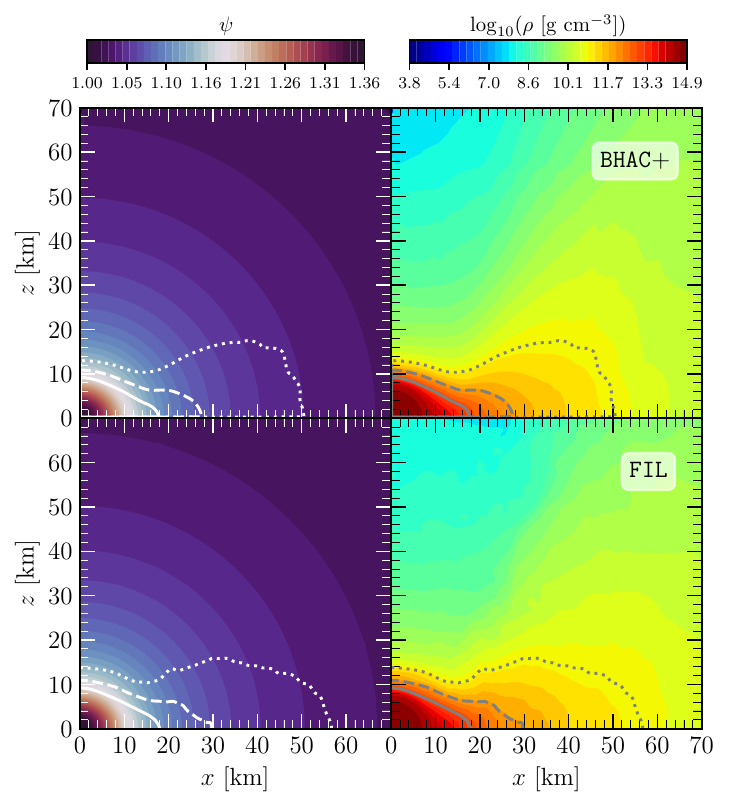}
\hspace{0.9cm}
\includegraphics[width=0.47\textwidth]{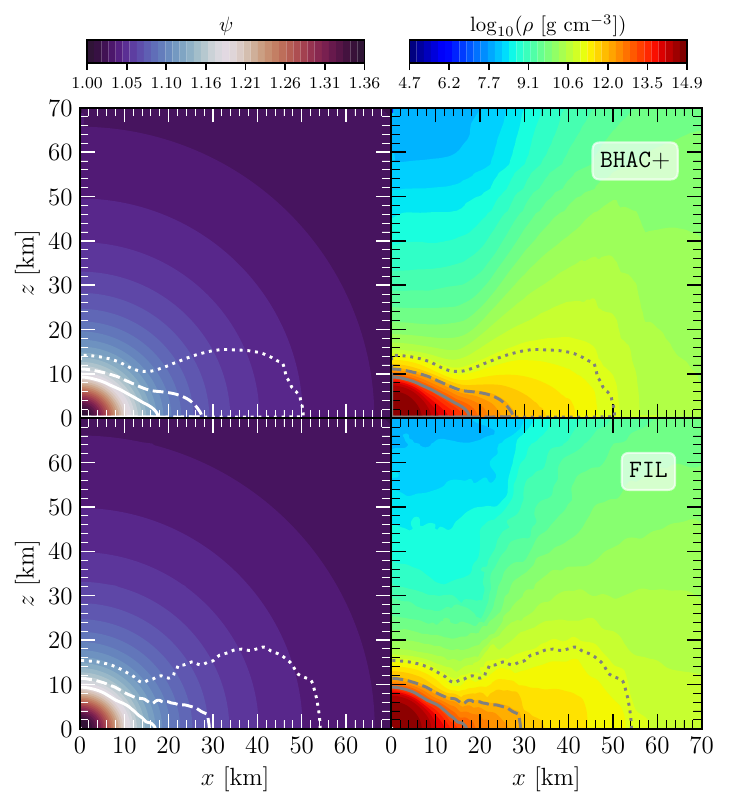}
\caption{\textit{Left panel:} 2D slices of the conformal factor (left)
  and of the rest-mass density (right) for a BNS post-merger remnant at
  $\bar{t} = 50\,\mathrm{ms}$ as evolved by \BHAC (top part of the panel)
  and by \FIL (bottom part); the data has been handed-off at
  $t_{\mathrm{HO},1} = 20\,\mathrm{ms}$. \textit{Right panel:} The same
  as on the left but at $\bar{t} = 100\,\mathrm{ms}$ and evolved with
  data handed-off at $t_{\mathrm{HO},2} = 50\,\mathrm{ms}$ (see
  Fig.~\ref{fig:DD2BNS_1dslices} for a quantitative comparison in 1D).
  The white (grey) solid, dashed and dotted contours in the left
    (right) part of each panel refer to rest-mass densities of $10^{13}
    \, \rm g/cm^3$ (solid lines), $10^{12} \, \rm g/cm^3$ (dashed lines),
    and $10^{11} \, \rm g/cm^3$ (dotted lines), respectively. The
    colormap for the conformal factor is tuned to highlight the location
    of rest-mass densities of the order of the $10^{13} \, \rm g/cm^3$,
    which may be taken as reference for the location of the surface of
    the HMNS.}
\label{fig:DD2BNS_2dslices_20msto50ms}
\end{figure*}

\subsection{Long-term evolution of a post-merger remnant}
\label{sec:longtermBNS}

The next and final test we present is related to the 2D long-term (\ie
for about one second) evolution of the remnant of a BNS
merger. Although the evolutions in \BHAC are only in 2D, this test
  is actually more challenging than the previous one in 3D, as it
  stress-tests the evolution on very long timescales, over which
  instabilities or numerical-dissipation effects may manifest.

A number of recent studies (see, \eg~\cite{Kastaun2014, Hanauske2016,
  Fujibayashi2017}) have shown that the BNS post-merger remnant relaxes
into a nearly axisymmetric and quasi-stationary state after a few tens of
milliseconds after the merger event. Furthermore, after approximately
$50\,\mathrm{ms}$ after the merger, the absolute magnitude of the
non-diagonal components of $\tilde{\gamma}_{ij}$ from \FIL become very
small, with relative differences with respect to the corresponding flat
components that is $\lesssim 2\%$. Under these conditions, a more
efficient and less expensive treatment of the spacetime evolution is
particularly useful, especially for long-term evolutions at high
resolutions. Of course, in order for \BHAC to perform such an evolution
it requires a consistent initial data and this can only be provided by a
full-numerical relativity code, such a \FILns, and the HO procedure
described in Sec.~\ref{sec:test_hand_off}.

In practice, after constructing the initial data for a binary system of
neutron stars with equal masses of $M = 1.40~\,{M_{\odot}}$ in
irrotational quasi-circular equilibrium and zero magnetic field with
\texttt{FUKA}, we evolve the system with \FIL well past the merger. The
evolution is handled using five levels of mesh refinement and with the
highest-resolution level having a spacing of $0.2\,M_{\odot}\simeq
0.295\,{\rm km}$. Defining $t$ and $t_\mathrm{mer}$ respectively as the
times since the start of the simulation and the merger time\footnote{As
customary, we define the merger time as the time of the global maximum of
the GW strain amplitude~\cite{Baiotti08}.}, we define the retarded time
as $\bar{t}_{\mathrm{HO}} := t - t_\mathrm{mer}$ and fix the HO time from
\FIL to \BHAC to a specific value of $\bar{t}$. Since the time of HO
represents an important (and to some extent arbitrary) aspect of the
long-term evolution, and in the spirit of assessing its impact, we have
carried out two distinct simulations with HO at $t_{\mathrm{HO},1} :=
20\,\mathrm{ms}$ and $t_{\mathrm{HO},2} := 50\,\mathrm{ms}$,
respectively.

\begin{figure*}
\includegraphics[width=0.90\textwidth]{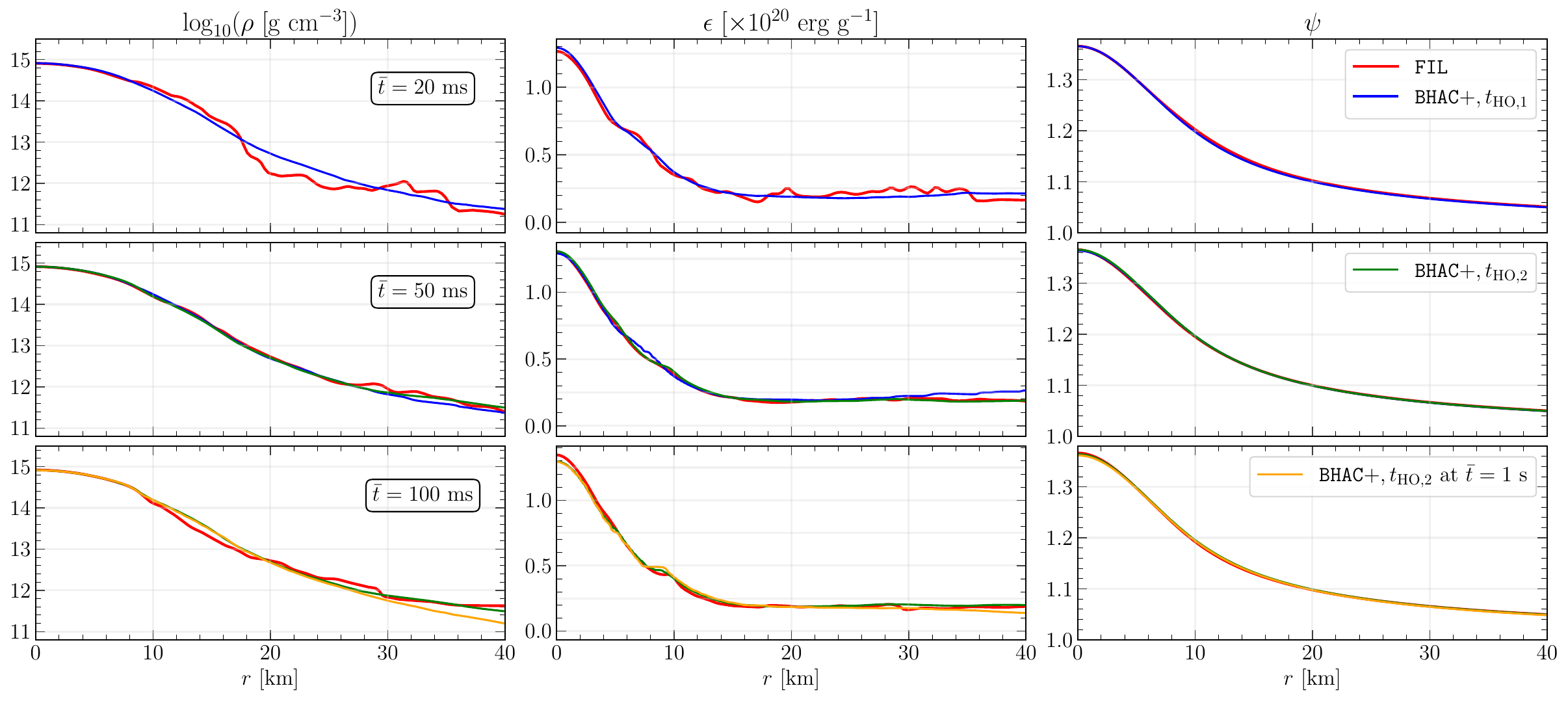}
\caption{1D slices at $z=0$ of the rest-mass density (left column), of
  the specific internal energy (middle column) and of the conformal
  factor (right column) for a BNS post-merger remnant at different times,
  \ie $\bar{t} = 20\,\mathrm{ms}$ (top row, test
  \texttt{DD2BNS-HO@20ms}), $\bar{t} = 50\,\mathrm{ms}$ (middle row, test
  \texttt{DD2BNS-HO@50ms}) and $\bar{t} = 100\,\mathrm{ms}$ (bottom row,
  test \texttt{DD2BNS-HO@50ms}). While the \FIL data is always indicated
  with a red line, the \BHAC data is shown with different colors
  depending on the HO time, \ie with a blue solid line in the top row for
  $t_{\mathrm{HO},1}$ and with a green solid line in the middle and
  bottom rows for $t_{\mathrm{HO},2}$. Also reported with an orange solid
  line in the bottom row is the \BHAC evolution at $t=1000\,{\rm ms}$
  (see Fig.~\ref{fig:DD2BNS_2dslices_20msto50ms} for a more qualitative
  comparison in 2D).}
\label{fig:DD2BNS_1dslices}
\end{figure*}

We start our comparison by showing in
Fig.~\ref{fig:DD2BNS_2dslices_20msto50ms} the 2D slices of the conformal
factor $\psi$ (left part of each panel) and of the rest-mass density
$\rho$ (right part of each panel) at two representative times, \ie
$\bar{t} = 50\,\mathrm{ms}$ (left panel) and $\bar{t} = 100\,\mathrm{ms}$
(right panel) respectively. Furthermore, for each panel, the top parts
report the solutions from \BHAC when the HO is made at
$t_{\mathrm{HO},1}$ (left panel) or $t_{\mathrm{HO},2}$ (right panel),
while the bottom part shows the solution relative to the \FIL evolution
restricted to the slice at $y=0$. Overall, the eight sub-panels shown in
Fig.~\ref{fig:DD2BNS_2dslices_20msto50ms} indicate that, at least
qualitatively, the solutions coming from the two codes are remarkably
similar despite the differences in the approaches for the evolution of
the spacetime and the different dimensionality (3D for \FIL and 2D for
\BHACns). Of course, there are two main reasons for this very good
match. First, the gravitational fields characterising the remnant are
comparatively weak and rather slow-varying, so that the CFC approximation
provides a very good description. Second, by the time the HO is made at
$t_{\mathrm{HO},1}$, the remnant is significantly axisymmetric, so that
the azimuthally averaged description of the remnant made by \BHAC matches
very well the fully-3D solution computed with \FILns.

Figure~\ref{fig:DD2BNS_1dslices} goes from the qualitative description of
Fig.~\ref{fig:DD2BNS_2dslices_20msto50ms} to a more quantitative one by
reporting the 1D slices at $z=0$ for the solution obtained by \BHAC and
by \FIL\ (in this case the data is extracted at $y = z = 0$; red solid
lines) and at three representative times, namely,
$\bar{t}=20\,\mathrm{ms}$ (top row), $\bar{t}=50\,\mathrm{ms}$ (middle
row), and $\bar{t}=100\,\mathrm{ms}$ (bottom row) and for three different
quantities, the rest-mass density (left column), the specific internal
energy (middle column), and the conformal factor (right column). Note
that in the case of the \BHAC evolutions, we distinguish the data coming
from $t_{\mathrm{HO},1}$ (blue solid lines in the top and middle rows)
from that obtained when the HO is instead done at $t_{\mathrm{HO},2}$
(green solid lines in the middle and bottom rows). Note also that
simulation by \FIL is carried out till $\bar{t}=100\,{\rm ms}$, while the
\BHAC simulation with $t_{\mathrm{HO},1}$ till $\bar{t}=50\,{\rm ms}$,
and that with $t_{\mathrm{HO},2}$ is performed till $\bar{t}=1.0\,{\rm
  s}$. Finally, shown instead with an orange solid line in the bottom
row is the solution from \BHAC at the final time of $1000\,{\rm ms}$.

Let us first compare the behaviour of the rest-mass density (left column)
in the three different snapshots. Overall, it is clear that \BHAC can
reproduce the structure of the binary merger remnant very well,
especially in the inner regions (\ie $r \lesssim 10\,{\rm km}$) and quite
independently of the HO time. Obviously, since the \BHAC simulations are
in 2D only, the corresponding profiles are smoother than those from \FIL
but the differences are apparent only when reported in a logarithmic
scale, as we do in Fig.~\ref{fig:DD2BNS_1dslices}. Note also that the
rest-mass profile in the remnant does not change considerably between
$\bar{t}=100\,{\rm ms}$ (which represents the last time of the solution
from \FILns) and $\bar{t}=1000\,{\rm ms}$, with the structure of the
remnant from \BHAC being only slightly more diffused than that from \FIL
(\cf different profiles from $10 \lesssim r \lesssim 20\,{\rm km}$).

Similar considerations apply also to the specific internal energy (middle
row), where the \BHAC solution with $t_{\mathrm{HO},2}$ (red solid line)
shows a better agreement with the reference \FIL solution as compared to
that with $t_{\mathrm{HO},1}$ (blue solid line). The evolution of the
temperature profile, on the other hand, can show more visible differences
and is more sensitive on the HO time (not shown in
Fig.~\ref{fig:DD2BNS_1dslices}). More specifically, the \BHAC solution
with $t_{\mathrm{HO},1}$ shows larger values of the temperature in the
region $5 \lesssim r \lesssim 20\,{\rm km}$, and smaller values in the
more internal regions of the remnant, \ie for $r \lesssim 5\,{\rm km}$;
furthermore the temperature profile in \BHAC in this inner core also
exhibits oscillations that have small amplitude and short wavelengths.
The origin of these differences can be attributed to three main
origins. First, the initialization of the CFC field variables on the
initial slice inevitably introduces fluctuations that are magnified in
the behaviour of the temperature. This is due to the fact that the
  gauges undergo a sudden change upon import and that the given HO data
  is not purely conformally flat. In turn, the metric initialization
  induces slight differences in the values of $\rho, Y_e, \epsilon$.
Second, the initial differences in $\epsilon$ and $Y_e$ between \FIL and
\BHAC are of the same order as those in $\rho$ and, especially after
metric initialization, such initial differences become larger when the
new constraints are satisfied. As a result, the accuracy of the
calculation of the temperature as a function $T = T(\rho, \epsilon, Y_e)$
-- which already suffers from a poor resolution of the tabulated EOS at
these regimes and from a high-power dependence of $T$ from $\epsilon$ in
regions of high rest-mass density -- is further affected. Third, the
small fluctuations in the temperature produced in the conversion from
$\epsilon$ to $T$ in the table are more easily averaged in a 3D
simulation (where every cell has six neighbours to average with) than in
a 2D simulation. Moreover, the timeslice where we perform HO of the
  3D \texttt{head-on} simulation is generated by \texttt{FUKA}~at $t=0$,
  is nearly conformally flat, and this drastically reduces the
  differences induced by differences in the gauges between \FIL and
  \BHAC.  Indeed, we observe that these oscillations are absent in the
3D \texttt{head-on} simulation presented in Sec.~\ref{sec:head-on} or
when evolving the post-merger data from \FIL in
3D~\cite{Jiang2024:inprep}.

Finally, the right column of Fig.~\ref{fig:DD2BNS_1dslices} reports the
profiles of the conformal factor following the same convention in terms
reported times and of HO times as in the left and middle columns. The
comparison in this case is even simpler to describe and it is clear that
the differences are very small for all the configurations
considered. More specifically, the largest absolute relative
differences in the rest-mass density [conformal factor] at
$\bar{t}=20\,\mathrm{ms}$ (\FIL vs \BHAC with $t_{\mathrm{HO},1}$),
$\bar{t}=50\,\mathrm{ms}$ (\FIL vs \BHAC with $t_{\mathrm{HO},1}$), and
$\bar{t}=100\,\mathrm{ms}$ (\FIL vs \BHAC with $t_{\mathrm{HO},2}$) are
respectively $1.18\%$ [$0.05\%$], $0.17\%$ [$0.13\%$], and $1.13\%$
[$0.20\%$]. Even when comparing the \FIL solution at $\bar{t}=100 \,
\mathrm{ms}$ with the corresponding \BHAC solution with
$t_{\mathrm{HO},2}$ and at time $\bar{t} = 1000 \, \mathrm{ms}$, the
relative difference in the rest-mass density is $1.39\%$ (similar
relative differences, \ie $0.37\%$ are measured for the conformal
factor). Adding the radiation-reaction terms in the CFC scheme that have
been here ignored can only further decrease the differences measured in
the two evolutions.

\begin{figure}
\includegraphics[width=0.49\textwidth]{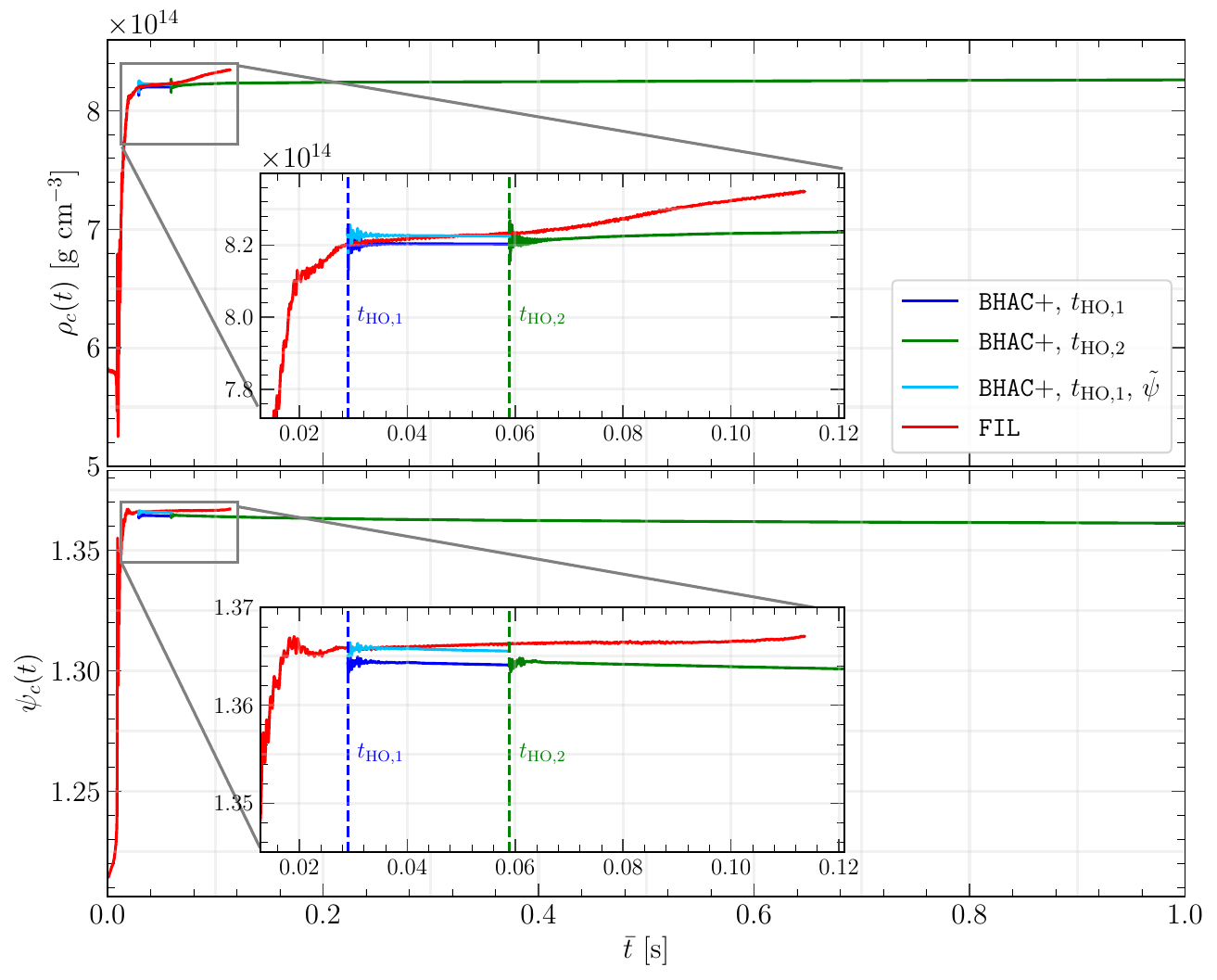}
\caption{Evolution of the central rest-mass density (top) and of central
  conformal factor (bottom) as obtained from different evolutions. In
  particular, the red solid line refers to the \FIL simulation carried
  out till $t=112\,\mathrm{ms}$, the blue solid line shows the \BHAC
  evolution with HO at $t_{\mathrm{HO},1}$, while the green solid line
  shows the \BHAC evolution with HO at $t_{\mathrm{HO},2}$ and continued
  till $t=1000\,\mathrm{ms}$. Also shown with a light-blue solid line is
  the \BHAC evolution with HO at $t_{\mathrm{HO},1}$ but with a rescaling
  of the conformal factor $\tilde{\psi} = 1.00038\, \psi$ to account for
  the slightly different spacetimes in \FIL and \BHACns.}
\label{fig:DD2BNS_rhopsic}
\end{figure}

We conclude this section on the long-term evolution of a post-merger
remnant by presenting in Fig.~\ref{fig:DD2BNS_rhopsic} a much more
precise comparison between the different evolutions. In particular, we
report in Fig.~\ref{fig:DD2BNS_rhopsic} the evolution of the maximum
values of the rest-mass density $\rho_c$ (top part) and of the
conformal factor $\psi_c$ (bottom part). Note that, as done so far in
other figures, we show with red solid lines the evolutions coming from
\FIL, while we indicate with either a blue or a green solid line the
evolutions from \BHAC with HOs at $t_{\mathrm{HO},1}$ and
$t_{\mathrm{HO},2}$, respectively. We should also remark that the
  main purpose of Fig.~\ref{fig:DD2BNS_rhopsic} is to show that the
  evolution in \BHAC does not suffer from stability problems and that,
  once provided with a matter configuration that is stable in the
  absence of gravitational radiation, it preserves this equilibrium
  for timescales that are $\sim 10$ times larger than those normally
  explored in full numerical-relativity codes. On the other hand,
  because gravitational radiation-reaction terms are neglected in
  \BHACns, the evolution over such long timescales can differ (even
  qualitatively) from that obtained with full numerical-relativity
  codes.

The first piece of information that can be readily appreciated from
Fig.~\ref{fig:DD2BNS_rhopsic} is that the differences in the evolution of
the two quantities are of the order of $\sim 1-2\%$ over the whole
timescale in which the \FIL evolution is carried out, \ie
$\bar{t}=100\,{\rm ms}$ (see insets). Hence, this figure provides a
strong and reassuring evidence that the use of the CFC approximation does
not yield to large quantitative differences in either the gravitational
fields or the matter variables even in the regions of strongest
curvature. The second piece of information is also quite self-evident:
the central rest-mass density and the conformal factor grow linearly with
time in the \FIL evolution, while they remain essentially constant in the
evolutions with \BHAC (in practice the central rest-mass density
increases of $+1.8\%$ from $\bar{t} = t_{\mathrm{HO},2}$ to
$1~\mathrm{s}$) This is not surprising and reflects the fact that the
emission of GWs in \FILns, and the consequent loss of energy and angular
momentum in the remnant, leads to an increase of its compactness and
hence of the central rest-mass density and conformal
factor~\cite{Baiotti08}. Since these losses are neglected in the present
implementation of the xCFC equations, the evolutions with \BHAC can only
show a slight increase in the central rest-mass density, as shown by blue
and green solid lines.

The third and final piece of information in Fig.~\ref{fig:DD2BNS_rhopsic}
comes from noting that while the differences in the evolutions with \FIL
and \BHAC are minute and much smaller than the uncertainties that
accompany the evolution of the post-merger remnant when all the physical
and microphysical effects are taken into account, these differences can
be reduced through a simple but artificial rescaling of the conformal
factor at $t_{\mathrm{HO},1}$. In particular, the inset in the bottom
panel of Fig.~\ref{fig:DD2BNS_rhopsic} reveals that the value of $\psi_c$
at HO drops by a factor $\sim 0.1\%$ (compare red and blue solid lines)
as a result of the mismatch between the different descriptions of the
spacetime in the two codes. This mismatch, however, can be easily
compensated by a global rescaling of the conformal factor by a constant
coefficient $1.00038$ and is shown by the evolutions indicated with
light-blue solid lines. When comparing these lines with the corresponding
evolutions without the rescaling (blue solid lines) it becomes apparent
that the match with \FIL can be easily improved, albeit rather
artificially. 

As a concluding remark, we note that while the differences in the \FIL
and \BHAC evolutions of the post-merger remnant are of the order of a
couple of percent, the corresponding computational costs differ by a
factor $\sim 65$. In particular, while the evolution in \FIL between the
HO time $t_{\mathrm{HO},2}$ and the end of the simulation at
$\bar{t}=100\,\mathrm{ms}$ implied a computational cost of
$\sim2.75\times10^5$ CPU hours, the same evolution with \BHAC incurred in
$\sim 4.25\times10^{3}$ CPU hours. Furthermore, the whole computational
cost in \BHAC from $t_{\mathrm{HO},2}$ till the end of the simulation at
$\bar{t}=1000\,\mathrm{ms}$ corresponded to $\sim 8.59\times10^4$ CPU
hours; assuming a stable remnant, an evolution to one second with \FIL
would have corresponded to a computational cost of $\sim 5.8\times10^6$
CPU hours, thus making it prohibitive if employed for a large number of
binaries.

\begin{figure}
\includegraphics[width=0.49\textwidth]{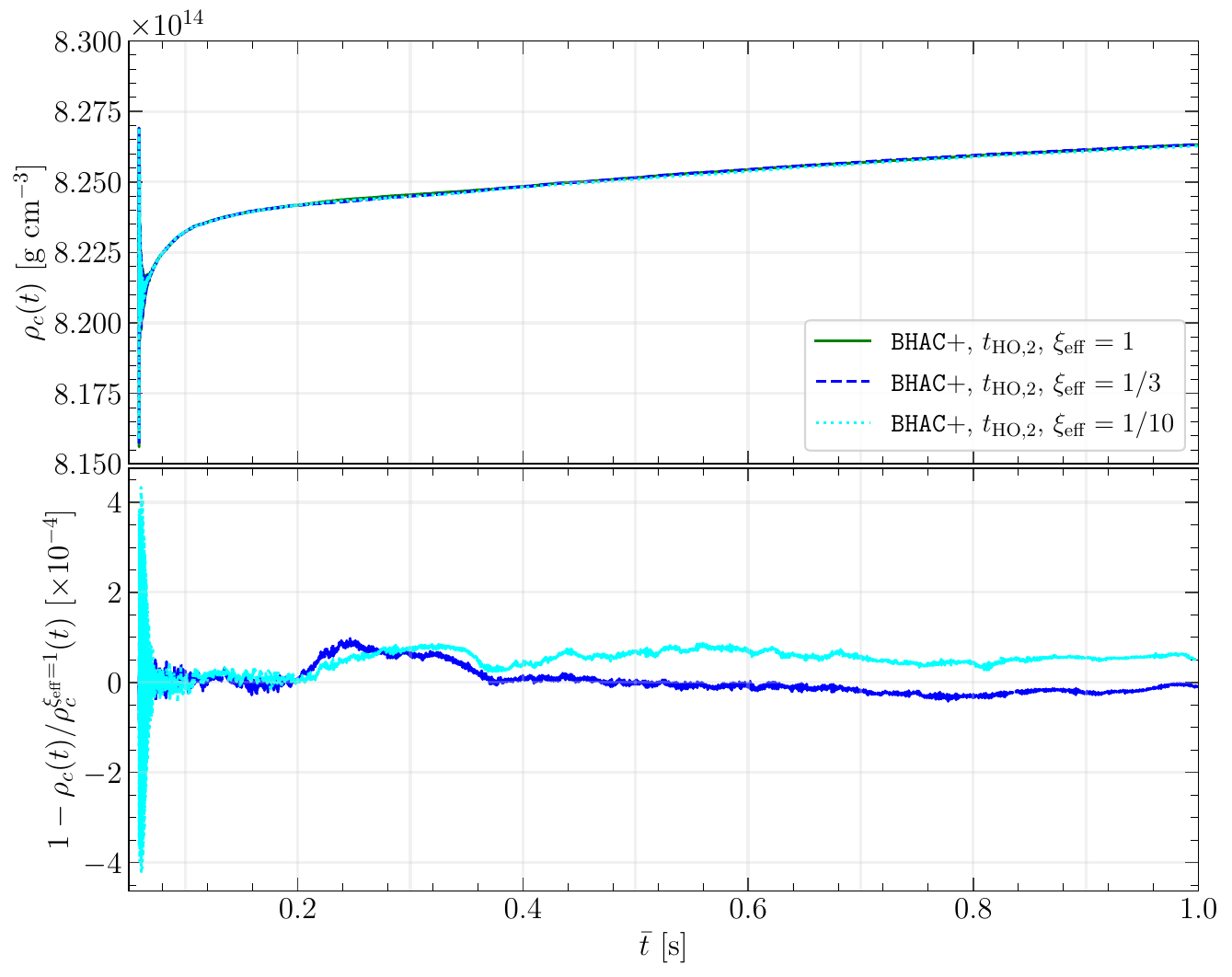}
\caption{\textit{Top panel:} Comparison of the evolutions of the central
  rest-mass density of the BNS remnant when different efficiency ratios
  are employed. In particular, the top panel reports the evolution with
  data at $t_{\mathrm{HO},2}$ when $\xi_{\rm eff}=1$ (green solid line;
  this is the same as in the top panel of Fig.~\ref{fig:DD2BNS_rhopsic}),
  $\xi_{\rm eff}=1/3$ (blue dashed line) and $\xi_{\rm eff}=1/10$ (cyan
  dotted line). The bottom panel reports instead the relative difference
  with respect to the $\xi_{\rm eff}=1$ reference evolution (blue and
  cyan solid lines for $\xi_{\rm eff}=1/3$ and $1/10$, respectively) and
  shows that the variance is $\lesssim 0.01\%$ even for the most extreme
  case of $\xi_{\rm eff}=1/10$; the latter simulation was performed with
  a computational gain of a factor $3.6$ with respect to the
  corresponding $\xi_{\rm eff}=1$ simulation.}
\label{fig:comp_costs}
\end{figure}

Finally, as mentioned in Sec.~\ref{sec:matter_solver}, the computational
costs relative to \BHAC can be further and easily be reduced by a factor
$2-3$ if less frequent solutions of the CFC equations are performed; in
the comparison presented above, in fact, we have solved the CFC equations
with the same frequency as the matter evolution and hence with a
spacetime slicing that is similar to that in \FILns. On the other hand,
this was not strictly necessary and the top panel of
Fig.~\ref{fig:comp_costs} reports the \BHAC post-merger evolution from
from $t_{\mathrm{HO},2}$ till the end of the simulation at $\bar{t} =
1000\,\mathrm{ms}$ when using $\xi_{\rm eff}=1, 1/3$ and $1/10$ (solid
green, dashed blue and dotted cyan lines, respectively). Clearly, the
three evolutions are extremely similar and the differences reported in
the bottom part of Fig.~\ref{fig:comp_costs} are always $\lesssim 0.01\%$
even in the extreme case of $\xi_{\rm eff}=1/10$. When exploiting this
additional speed-up, it is clear that the computational costs at this
resolution can be decreased by a factor $2.1$ and $3.6$ for $\xi_{\rm
  eff} = 1/3$ and $1/10$, respectively. As a result, the 2D \BHAC
simulation with $\xi_{\rm eff}=1/10$ has an effective gain over the
corresponding 3D simulation in \FIL of a factor $\simeq 230$. This
computational saving can only increase when considering higher
resolutions and opens the way to the systematic study in 2D of the
secular matter and electromagnetic emission from binary-merger remnants.

\section{Conclusions and Outlook}
\label{sec:summary}

One of the main challenges to be faced when modelling BNS mergers is the
accurate long-term evolution of the post-merger remnant over timescales
of the order of several seconds. When this modeling is made including all
the relevant aspects of the complex physics accompanying the remnant --
and which includes the proper treatment of magnetic fields, of realistic
EOSs, and of neutrino transport -- the computational costs can easily
become enormous. To address this challenge in part, we have developed a
novel hybrid approach that couples the full numerical-relativity GRMHD
code \FILns~\cite{Most2019b} with the versatile, multi-coordinate and
multi-dimensional GRMHD code~\texttt{BHAC}, which possesses robust
divergence-cleaning methods~\cite{Porth2017} and constraint-transport
methods~\cite{Olivares2019} for the enforcement of the divergence-free
condition of the magnetic field. However, because~\texttt{BHAC} was
developed to solve the equations of GRMHD on arbitrarily curved but fixed
spacetimes, we have extended the code capabilities by constructing
\BHACns, which employs the CFC approximation of the Einstein
equations. More specifically, by assuming a locally flat conformal
metric, such an approximation simplifies the Einstein equations reducing
them to a set of elliptic equations that can be solved to compute the
evolution of the spacetime as a response to the changes in the
energy-momentum tensor. A number of applications in core-collapse
simulations, but also in the study of merging neutron-star binaries, have
shown that the CFC approximation achieves good agreement with full
general-relativistic simulations, especially in isolated systems with
axisymmetry. The most important advantage of the CFC approximation is
however that the corresponding elliptic equations need to solved only
every $3-100$ steps of the underlying
hydrodynamical/magnetohydrodynamical evolution, thus allowing to the
capture even the highest-frequency modes of a fluid compact object at a
fraction of the computational cost.

We have therefore presented in detail the basic features of the new code
\BHACns, illustrating both the numerical setup for the solution of the
Einstein and GRMHD equations, and the strategies necessary to interface
\BHAC with a fully general-relativistic code, such as \FILns, when
importing both 2D and 3D data. Furthermore, we have methodically
described our implementation of an efficient and reliable
primitive-recovery scheme coupled with a finite-temperature and tabulated
EOS, demonstrating not only its robustness under a large variety of
physical regimes, but also its efficiency, which is comparable to (if not
higher than) the best reported in the literature so far.

In addition to describing our new methodology, we have also shown the
results of a series of standard and non-standard benchmark tests that
have been carried out to validate the various parts of the code. These
tests have been carried out with various coordinates systems and
different numbers of spatial dimensions, from 1D to fully 3D simulations,
and for timescales ranging from $5$ to $1000$ milliseconds. More
specifically, our tests have considered the simulation of oscillating
spherical stars with either a fixed or dynamical spacetime, the dynamics
of an unstable spherical star migrating over to the branch of stable
configurations, the simulation of a differentially rotating star endowed
with a strong toroidal magnetic field, as well as the long-term stability
of an unmagnetized but rapidly rotating star near the mass-shedding
limit. In many of these tests, the evolution carried out with \BHAC has
been compared with the equivalent one carried out with \FILns, finding
always a very good agreement.

Our list of benchmarks has been completed by two additional and
challenging tests. The first one has involved the head-on collision of
two equal-mass stars obeying a temperature-dependent EOS, whose dynamics
has been compared in detail with the corresponding one obtained with \FIL
revealing a remarkable agreement despite the very different handling of
the spacetime. The second one, instead, has explored in the detail the
hand-off procedure of data from \FIL to \BHAC in the 2D long-term
evolution of a BNS merger remnant. In particular, we have demonstrated
\BHACns's ability of importing -- even at different times -- matter and
spacetime data from \FIL describing the merger remnant and further
evolving it for hundreds of milliseconds and up to one second after
merger. While the evolution with \BHAC cannot by construction reproduce
the small changes observed in \FILns's remnant as a result of the
emission of GWs, the agreement between the two evolutions is very good
and below a couple of percent. More importantly, the computational costs
between the 3D \FIL evolution and the 2D \BHAC evolution differ by a
factor up to $230$ even at the modest resolutions considered here, with
the computational gain becoming larger as the resolution is
increased. This substantial difference opens the way to a much more
systematic exploration of the long-term evolution of post-merger
remnants, that is no longer restricted to considering a single specific
case, but can include variations in mass, EOS, magnetic-field properties,
etc.

Overall, the results presented here provide two main evidences. First,
\BHAC is able to accurately reproduce the evolution of compact objects in
non-vacuum spacetimes and the use of the CFC approximation reproduces
accurately both the gravitational fields and the matter variables even in
the regions of strongest curvature. Second, a hybrid approach in which a
short-term but full numerical-relativity treatment of the dynamics of
merging binaries is coupled to a long-term but CFC approximation for the
evolution of the post-merger remnant has great potential to obtain an
accurate description of the secular electromagnetic and matter emission
from binary mergers.

At the same time, a number of improvements can to be implemented to
achieve an even more accurate and physically realistic description of
these scenarios. These include the addition of radiation-reaction terms
in the CFC approximation, of the coupling of the GRMHD equations with
those of neutrino radiative transfer, and the handling of scenarios of
black-hole formation. We will report about these improvements in future
works.

\begin{acknowledgments}
  It is a pleasure to thank Patrick Cheong Chi-Kit, Alan Lam Tsz-Lok,
  Hector Olivares and Oliver Porth for the helpful and detailed
  discussions on implementations. We also thank Frederike Kubandt for her
  initial work on the hand-off between \FIL and \BHACns. This research is
  supported by the ERC Advanced Grant ``JETSET: Launching, propagation
  and emission of relativistic jets from binary mergers and across mass
  scales'' (grant No. 884631), by the Deutsche Forschungsgemeinschaft
  (DFG, German Research Foundation) through the CRC-TR 211
  ``Strong-interaction matter under extreme conditions'' -- project
  number 315477589 -- TRR 211. J.L.J. acknowledges support by the
  Alexander von Humboldt Foundation.

\paragraph{Software:}
\texttt{BHAC}~\cite{Porth2017,Olivares2019,Ripperda2019},
\texttt{FIL}~\cite{Most2019b},
\texttt{ETK}~\cite{Loffler:2011ay},
\texttt{RNS}~\cite{Stergioulas95},
\texttt{FUKA}~\cite{Papenfort2021},
\texttt{XNS}~\cite{Bucciantini2011,Pili2014},
\texttt{StellarCollapse} (\href{https://stellarcollapse.org}{https://stellarcollapse.org}), 
\texttt{CompOSE} (\href{https://compose.obspm.fr}{https://compose.obspm.fr})
\end{acknowledgments}

\appendix

\section{On the coordinate transformations between \FIL and \BHAC}

The HO procedure of data from \FIL to \BHAC presented in the main
  text requires the transformation of quantities expressed in Cartesian
  coordinates $x^\mu=\{t,x,y,z\}$ within \FIL to cylindrical coordinates
  $\bar{x}^\mu=\{t,r,\phi,z\}$ within in \BHACns. While the relevant
  transformations between these sets of coordinates is elementary, we
  repeat it here for completeness.\\
  Since the transformation is performed on a fixed timeslice (\ie $t=\rm
  const.$) it is sufficient to consider here only the transformation laws
  of the spatial components.  Furthermore, the $z$-coordinate, which
  coincides with the rotation axis in the cylindrical coordinate system,
  does not change when going from Cartesian to cylindrical coordinates
  and vice versa. This leaves the following, non-trivial relations that
  implement the coordinate change. More specifically, the Cartesian $x$
  and $y$ coordinates can be written in terms to the cylindrical radius
  $r$ and angle coordinate $\phi$ at fixed $z$ as
  \begin{equation}
    x(r,\phi)=r\cos(\phi)\,,\qquad y(r,\phi)=r\sin(\phi)\,.
  \end{equation}
  The transformations of the metric and of a generic vector field
    $\boldsymbol{B}$, \eg the magnetic field, components follow then from
    the usual tensor transformation laws
  \begin{equation}
    \bar{g}_{\mu\nu}(\bar{x}^\delta)=\frac{\partial x^\rho}{\partial \bar{x}^\mu}\frac{\partial x^\sigma}{\partial \bar{x}^\nu}g_{\rho\sigma}(x^\delta)\,,\quad \bar{B}^{\mu}(\bar{x}^\delta)=\frac{\partial \bar{x}^\mu}{\partial x^\rho} B^\rho(x^\delta)\,,
  \end{equation}
  where the Jacobian and its inverse are given by
  \begin{eqnarray}
    J^\mu_\nu&=&\frac{\partial x^\mu}{\partial \bar{x}^\nu}=
    \begin{pmatrix}
      \cos(\phi)& -r \sin(\phi)& 0\\
      \sin(\phi)&  r \cos(\phi)& 0\\
      0  &      0       & 1
    \end{pmatrix}\,,\\
    \nonumber \\
    J^\nu_\mu&=&\frac{\partial \bar{x}^\nu}{\partial x^\mu}=
    \begin{pmatrix}
      \cos(\phi)&  \sin(\phi)& 0\\
      -\sin(\phi)/r&  \cos(\phi)/r& 0\\
      0  &      0       & 1
    \end{pmatrix}\,.
  \end{eqnarray}
  The spatial components of the Cartesian metric $g_{ij}$ and of the
    vector field $B^i$ can then be expressed in terms of cylindrical
    coordinates as follows
  \begin{eqnarray}
    &&\bar{g}_{rr}=\cos(\phi)^2 g_{xx}+\sin(2\phi)g_{xy}+\sin(\phi)^2 g_{yy}\,,\\
    &&\bar{g}_{r\phi}= r \cos(2\phi)g_{xy}+r\sin(\phi)\cos(\phi)(g_{yy}-g_{xx})\,,\\
    &&\bar{g}_{rz}= \cos(\phi)g_{xz}+\sin(\phi)g_{yz}\,,\\
    &&\bar{g}_{\phi\phi}= r^2\left(\cos(\phi)^2g_{yy}-2\sin(\phi)\cos(\phi)g_{xy}
    +\sin(\phi)^2g_{xx}\right)\,,\nonumber\\ \\
    &&\bar{g}_{\phi z}=r\left(\cos(\phi)g_{yz}-\sin(\phi) g_{xz} \right)\,,\\
    &&\bar{g}_{z z}=g_{zz}\,,
  \end{eqnarray}
  and
  \begin{eqnarray}
    \bar{B}^r&=&\cos(\phi)B^x+\sin(\phi)B^y\,,\\
    \bar{B}^\phi&=&\frac{\cos(\phi)B^y-\sin(\phi)B^x}{r}\,,\\
    \bar{B}^z&=&B^z\,.
  \end{eqnarray}

  In general, when handing over variables from one code to the
    other, the location of the gridpoints in the two coordinate systems
    do not coincide and need to be interpolated, which we do with a
    standard first-order interpolation. Furthermore, since \FIL is a
    higher-than-second order accurate finite difference code, the
    interpolations are performed on point-wise values of both metric and
    hydrodynamic variables at the gridpoint locations.

\bibliography{aeireferences}

\end{document}